\documentclass[english,aps,prd,a4paper,groupedaddress,preprintnumbers,floatfix,nofootinbib,12pt]{revtex4}

\usepackage{graphicx}
\usepackage{enumerate}
\usepackage{amsmath}
\usepackage{amssymb}
\usepackage{amsfonts}
\usepackage{xcolor}
\usepackage{url}
\usepackage{hyperref}
\usepackage{booktabs}
\usepackage{caption}
\usepackage[utf8]{inputenc}

\usepackage{arydshln}
\usepackage{natbib}

\hypersetup{colorlinks=true,linkcolor=redLinks,citecolor=greenLinks,
urlcolor=redLinks,
pdfborder={0 0 1}}
\hypersetup{
    bookmarks=true,         
    unicode=false,          
    pdftoolbar=true,        
    pdfmenubar=true,        
    pdffitwindow=false,     
    pdfstartview={FitH},    
    pdftitle={My title},    
    pdfauthor={Author},     
    pdfsubject={Subject},   
    pdfcreator={Creator},   
    pdfproducer={Producer}, 
    pdfkeywords={keyword1, key2, key3}, 
    pdfnewwindow=true,      
    colorlinks=false,       
    linkcolor=red,          
    citecolor=green,        
    filecolor=magenta,      
    urlcolor=cyan           
}

\newcommand{\bi}{\begin{itemize}}
\newcommand{\ei}{\end{itemize}}

\begin{document}
\title{Coherent elastic neutrino-nucleus scattering as a precision test for 
the Standard Model and beyond: the COHERENT proposal case}
\author{O. G. Miranda$^1$} \email{omr@fis.cinvestav.mx}
\author{G. Sanchez Garcia$^1$} \email{gsanchez@fis.cinvestav.mx}
\author{O. Sanders$^1$} \email{osanders@fis.cinvestav.mx}
\affiliation{$^1$~Departamento de F\'isica, Centro de Investigaci\'on
  y de Estudios Avanzados del IPN, Apdo. Postal 14-740, 07000 Ciudad
  de M\'exico, M\'exico.}    

\begin{abstract}\noindent
Several experimental proposals expect to confirm the recent
measurement of the coherent elastic neutrino-nucleus scattering
(CEvNS). Motivated in particular by the next generation experiments of
the COHERENT collaboration, we study their sensitivity to different
tests of the Standard Model and beyond. We analyze the resolution that
can be achieved by each future proposed detector in the measurement of
the weak mixing angle; we also perform a similar analysis in the
context of Non-Standard Interaction (NSI) and in the case of
oscillations into a sterile neutrino state.  We show that future
perspectives are interesting for these types of new physics searches.
\end{abstract}

\maketitle

\section{Introduction}
Despite the coherent elastic neutrino-nucleus scattering (CEvNS) was proposed 
more than forty years ago~\cite{Freedman:1973yd}, it was only recently that 
the COHERENT collaboration observed
this process for the first time by using a CsI[Na] detector
exposed to the neutrino flux generated at the Spallation Neutron Source
(SNS) at Oak Ridge National Laboratory~\cite{Akimov:2017ade}.

In a CEvNS process, an incident neutrino interacts coherently with the
protons and neutrons within the nucleus. As a result, there is an
enhancement in the cross-section, which turns out to be quadratic in
the number of nucleons.  The necessary condition to observe this
phenomenon is that the energy of the neutrino must be sufficiently low
so that the momentum transfer satisfies $qR << 1$, with $R$ the
nuclear radius.  Since its first detection, COHERENT data have been
studied for different purposes such as measurements of nuclear neutron
distributions~\cite{Cadeddu:2017etk}, weak mixing
angle~\cite{Cadeddu:2018izq,Huang:2019ene}, neutrino electromagnetic
properties~\cite{Kosmas:2017tsq,Cadeddu:2018dux}, and tests of NSI
neutrino
interactions~\cite{Coloma:2017ncl,Liao:2017uzy,Kosmas:2017tsq}.

In the future, the COHERENT program~\cite{Akimov:2018ghi} will include
a set of four detectors, each based on different materials and
technologies capable of observing low-energy nuclear recoils: the
currently used CsI[Na] scintillating crystal, with which CEvNS was
detected for the first time, and three future experiments that are
still being developed: a set of p-type point-contact Germanium
detectors, a single-phase liquid Argon detector, and an array of
NaI[Tl] crystals. Each detector has a different threshold, baseline,
and mass, all of which are summarized in Table~\ref{Tab:01}. In this
work we study the future experimental setups proposed by the COHERENT
collaboration in order to test the sensitivity of CEvNS to the weak
mixing angle and the search of new physics by two different
mechanisms; the first one through the introduction of parameters which
describe NSI and the other by introducing the possibility of a
specific neutrino flavor to oscillate into a sterile one. The same
original experimental proposal~\cite{Akimov:2018ghi} introduces a
discussion about non-standard interactions (NSI) as well as
implications for dark matter. In this work, for the NSI analysis, we
study both non-universal and flavor changing NSI parameters. Different
authors have already studied part of the potential of these detectors
in a different
context~\cite{Denton:2018xmq,Billard:2018jnl,Brdar:2018qqj,Altmannshofer:2018xyo,AristizabalSierra:2019zmy,Blanco:2019vyp}. Here
we focus on the specific configurations reported by the COHERENT
collaboration for its future stages~\cite{Akimov:2018ghi} to have a
complementary forecast that includes cases that have not been covered,
such as the future perspectives for the measurement of a weak mixing
angle for these detectors.

\begin{table}
  \begin{tabular}{l c c c c c c c  } \hline \hline
& $T_{thres}$ & Baseline  & Det. Tec. & Fid. Mass   \\ \hline \hline
$^{133}$Cs$^{127}$I 
& $5$~keV   &$19.3$~m & Scintillator & $14.6$~kg  &  \\ 
$^{72}$Ge 
&  $5$~keV  &$22$~m   & HPGe PPC& $10$~kg  & \\
$^{23}$Na$^{127}$I 
& $13$~keV  & $28$~m & Scintillator & $2000$~kg &  \\
$^{40}$Ar 
& $20$~keV  & $29$~m & Liquid scintillator & $1000$~kg &  \\ 
\hline \hline
  \end{tabular}\caption{\label{Tab:01} {Current and future experimental setups 
for the COHERENT collaboration detectors~\cite{Akimov:2018ghi}.  } }
\end{table}

\section{Coherent elastic neutrino-nucleus scattering}
Before discussing the future perspectives for CEvNS in a specific SNS
experiment, we present in this section the main characteristics of the
neutrino flux, cross section and form factors involved in the
prediction of the number of events measured by a given detector. The
neutrino beam used by the COHERENT collaboration consists of $\nu_e,
\nu_\mu $ and $\bar{\nu}_\mu$ fluxes coming from the SNS. These
neutrinos are produced by the $\pi^+$ decay-at-rest in the form
$\pi^+\rightarrow \mu^+ \nu_\mu$ and thus we have a mono-energetic
beam of muon neutrinos, known as "prompt" neutrinos, which can be
described by:

\begin{equation}
\frac{\mathrm{dN_{\nu _{\mu }}} }{\mathrm{d}E }=\eta\delta\left ( E-\frac{m_{\pi }^{2}-m_{\mu }^{2}}{2m_{\pi }} \right ).
\label{FluxDelta}
\end{equation}
\newline
Eventually, the $\mu^+$'s also decay to produce anti-muon neutrinos
and electron neutrinos, which together are known as "delayed neutrinos", and
which can be modeled, for energies up to 52.8 MeV,
as~\cite{Cadeddu:2017etk}:

\begin{equation}
\frac{\mathrm{dN_{\overline{\nu} _{\mu }}} }{\mathrm{d}E }= \eta \frac{64E^{2}}{m_{\mu }^{3}}\left ( \frac{3}{4}-\frac{E}{m_{\mu }} \right )
\label{FluxMuon}
\end{equation}
\newline
\begin{equation}
\frac{\mathrm{dN_{\nu _{e }}} }{\mathrm{d}E }= \eta \frac{192E^{2}}{m_{\mu }^{3}}\left ( \frac{1}{2}-\frac{E}{m_{\mu }} \right )
\label{FluxElectron}
\end{equation}
\newline
Being $\eta = rN_{POT}/4\pi L^{2}$ a normalization factor with $r =
0.08$ the number of neutrinos per flavor, $N_{POT} = 1.76 \times
10^{23}$, the number of protons on target, and $L$ the distance
between the source and the detector. The total neutrino flux is
considered to be the sum of the three previous contributions. For all
our computations we will consider the same total flux, and we set it
as equal to that of the first COHERENT
measurement~\cite{Akimov:2017ade}; in this way, our comparison of
results will be made using the same standard time window.

Regarding the CEvNS cross section, this has been computed to be~\cite{Drukier:1983gj,Barranco:2005yy,Patton:2012jr,Papoulias:2015vxa}: 

\begin{equation} 
\left(\frac{d\sigma}{dT}\right)_{\rm SM}^{\rm coh} = \frac{G_{F}^{2}M}{\pi}\left[1-\frac{MT}{2E_{\nu}^{2}}\right]
 [Zg^{p}_{V}F_Z(q^2)+Ng^{n}_{V}F_N(q^2)]^{2}. \label{eq:00}
\end{equation}
\newline
\noindent Here, $M$ is the mass of the nucleus, $E_{\nu}$ is the
neutrino energy, and $T$ is the nucleus recoil energy; $F_{Z,N}(q^2)$
are the corresponding nuclear form factors, which are especially
important at higher momentum transfer, as is the case of neutrinos
coming from the SNS. In other cases, as for antineutrinos coming from
nuclear reactors, these form factors have a minimal impact due to the
low momentum transfer.  We have computed our results by using a Helm
form factor as well as a symmetrized Fermi one for both protons and
neutrons; our results in all cases were the same up to the level of
one per thousand, so in what follows, we will consider the Helm form
factor for neutrons and the symmetrized Fermi one for protons.  The
neutral current vector couplings are given by:
\begin{eqnarray}
\nonumber g_{V}^{p} &=&\frac{1}{2}-2\hat{s}_{Z}^{2}
\\ 
g_{V}^{n} &=&-\frac{1}{2}
\end{eqnarray}
\noindent where $\hat{s}_{Z}^{2}=\sin^2\theta_{W}=0.23865$, which corresponds to the low energy limit as well.~\cite{Patrignani:2016xqp}. 
%

\noindent Recently, a new computation that studies in more detail the
cross section for the case of a non-zero spin nucleus (taking into
account the kinematics for relatively high momentum transfer) has been
reported~\cite{Bednyakov:2018mjd}. It has been stated that kinematic
corrections could be important, while axial couplings due to the
nuclear spin have less impact. In this picture, the CEvNS cross section
is given by~\cite{Bednyakov:2018mjd}:
\newline
\begin{equation}
\begin{aligned}
\frac{\mathrm{d} \sigma}{\mathrm{d} T} = \frac{G_{F}^{2}M}{\pi }g_{c}&\left ( 1 - \frac{TM}{2E_{\nu }} \right )\sum_{f,f'}F_{f}F_{f'}^{*}\left [ g_{V}^{f}g_{V}^{f'}\left ( A_{f}A_{f'}\left ( 1 - \frac{y\tau }{2} \right )^{2}+ \Delta A_{f}\Delta A_{f'}\left ( \frac{y}{2} \right )^{2} \right )\right. \\
&\left. + g_{A}^{f}g_{A}^{f'}\left ( A_{f}A_{f'}\left ( \frac{y\tau }{2} \right )^{2}+ \Delta A_{f}\Delta A_{f'}\left (1- \frac{y}{2} \right )^{2} \right )\right.\\
&\left. + 2g_{V}^{f}g_{A}^{f'}\left ( A_{f}A_{f'}\left ( 1 - \frac{y\tau }{2} \right )\frac{y\tau }{2}+ \Delta A_{f}\Delta A_{f'}\frac{y}{2}\left (1- \frac{y}{2} \right ) \right )\right ] 
\label{eq:BN}
\end{aligned}
\end{equation}
\newline
Where the sums on both $f$ and $f'$ run over $p$ and $n$, with $A_{p}
= Z$, $A_{n} = N$, and $\Delta A_{f}$ is the difference between the
corresponding nucleons with a spin projection along the incident
neutrino axis and those with spin projection opposite to it, the
Bjorken $y$ is given by $y = T/E_{\nu}$ and $s$ is the total energy
squared.  Finally, $\tau=\frac{\sqrt{s}-m_N}{\sqrt{s}+m_N}$, with
$m_N$ the nucleon mass. It has also been discussed in the same
reference that the contributions due to $\Delta A_{f}$ and
$g_{A}^{f}g_{A}^{f'}$ are small. We have checked that indeed, for
  the CsI cross section, this corrections are at least three orders of
  magnitude smaller and, therefore, we will not consider them.  After
these approximations, Eqs.~(\ref{eq:00}) and~(\ref{eq:BN}) are still
different by a factor $g_{c}$.  This factor arises if we
require~\cite{Bednyakov:2018mjd} that the interaction of the incident
neutrino happens only when the nucleon has an initial momentum
$\vec{p} = - (\vec{q}/2 )(1 - m_{N}/M)$, and acquires a final momentum
$\vec{p} + \vec{q}$, with $m_{N}$ the mass of the nucleon. In this
picture, the factor $g_{c}$ is given by the product of three different
factors, two of which are of order unity, while the last one is
reported to be linear in $T$~\cite{Bednyakov:2018mjd}. Under this
assumption, we found the factor $g_{c}$ to be given by:

\begin{equation}
g_{c} = 1 + \frac{MT}{m_{N}E_{\nu}} . \label{kinFactor}  
\end{equation}
\newline
On the other hand, once we take an expression for the cross
   section, the number of events measured by a detector is
   given by: 
\newline
\begin{equation}
N^{th} = N_{D}\int_{T}A(T)dT\int_{E_{min}}^{52.8 MeV}dE\lambda(E_{\nu},T)\frac{\mathrm{d} \sigma }{\mathrm{d} T},
\label{SM_Events}
\end{equation}
\newline
\noindent where $A(T)$ is an acceptance function, $\lambda (E_{\nu},
T)$ is the neutrino flux, and $N_{D}$ is, depending on the
detector, the number of targets in it and is given by
$N_{A}M_{det}/M_{D}$, with $N_{A}$ the Avogadro's number, $M_{det}$
the mass of the detector, and $M_{D}$ its molar mass. The limits of the
$T$ integral depend on both the detector's threshold and the
maximum recoil energy for a fixed $E_{\nu}$, which to our purposes is
well approximated by $T_{\rm max}(E_{\nu})\simeq 2E_{\nu}^{2}/M$. On
the other hand, the integral over $E_{\nu}$ has an upper limit of 52.8
MeV, which corresponds to the maximum energy of the neutrinos coming
from the SNS.

\noindent 
Before computing a forecast of the sensitivity to future new
experiments, we have computed what would be the expected number of
events in the case of the recent COHERENT detection of CEvNS for the
previous two formulations of the cross-section.  To this purpose, by
closely following the procedure described in
Ref.~\cite{Cadeddu:2017etk}, we have computed the expected number of
events by recoil energy bins for the case of the CsI detector using an
average neutron rms radius of $5.5$~fm for both Cs and I, which was
found to be the best fit to the COHERENT data~\cite{Cadeddu:2017etk};
we take the acceptance function as given
in~\cite{Akimov:2018vzs}. Table~\ref{Tab:01} (see
  Introduction) shows the specific values for the detector's mass and
its distance to the neutrino source.

\begin{figure}[ht] 
\begin{center}
\includegraphics[width=0.45\textwidth]{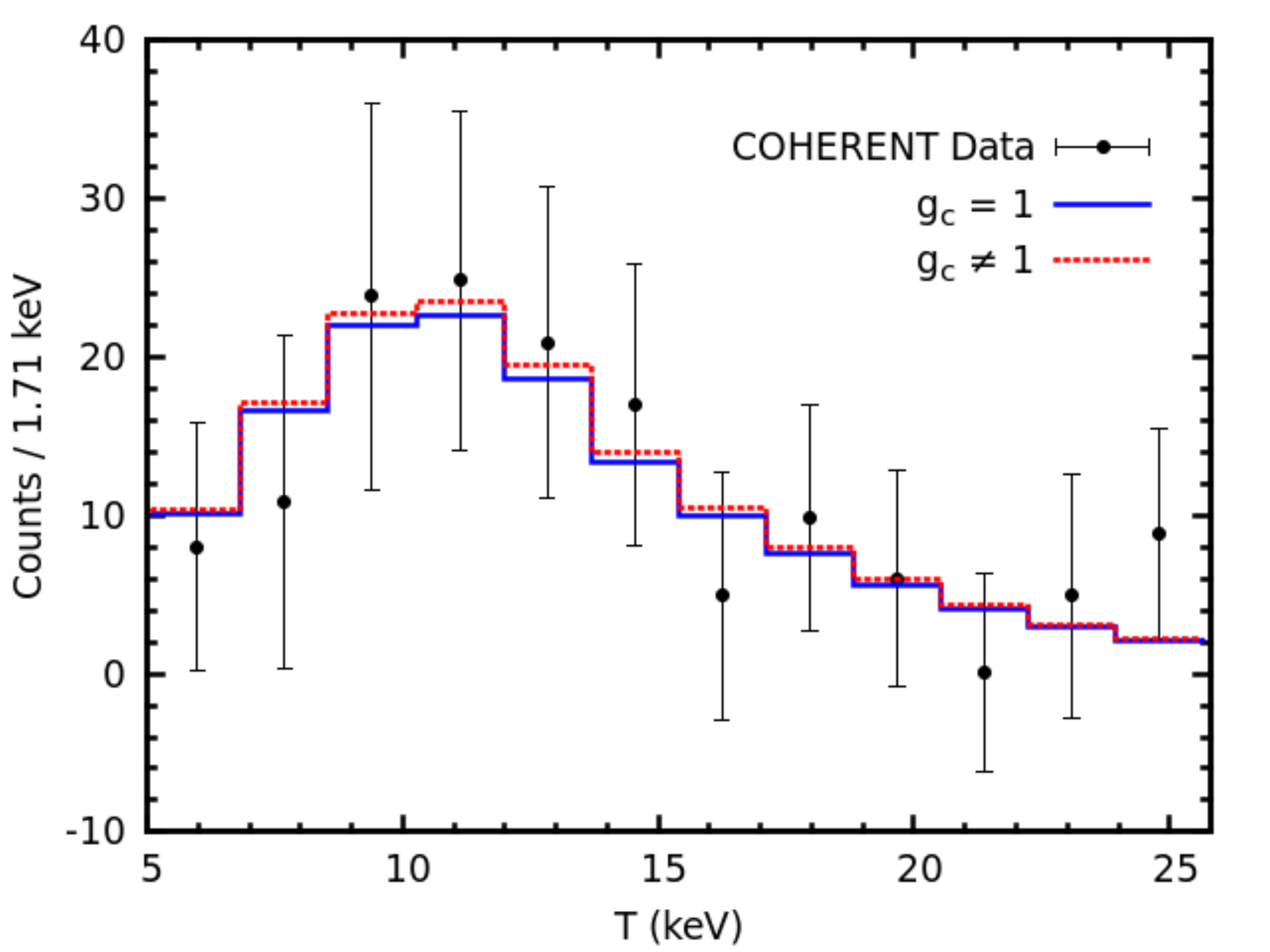}
\end{center}
\caption{\label{fig:comparison1} Expected number of events for the
  CsI COHERENT case. The solid (blue) line corresponds to the usual
  approach to the cross section as given in Eq.~(\ref{eq:00}) while
  the dotted (red) line corresponds to the more detailed case
  discussed in Ref.~\cite{Bednyakov:2018mjd}. The points correspond to 
  the experimental data~\cite{Akimov:2017ade}. }
\end{figure}

\noindent The results of this computation are illustrated in
Fig.~\ref{fig:comparison1}, where we show the expected number of
events when we consider the cross-section as in Eq.~(\ref{eq:00}) as
well as when we consider the case of a linear kinematic correction due
to the factor $g_{c}$. It is possible to notice that the introduction
of the kinetic factor $g_{c}$ yields to a relatively larger number of
events.  Despite this factor can introduce important corrections in
the cross section for high neutrino energies, the convolution of the
neutrino energy spectrum and form factors translates into an increase
of $5$~\% or less in the total number of events for the targets under
consideration.  We have checked that this effect translates into a
small shift in the central value of a given fit, but has no impact in
the width of the errors. Therefore, for our computations of the future
expectations we will show the results obtained with the more simple
and usual approach of Eq.~(\ref{eq:00}).

Regarding future experiments, throughout the following sections we
will study the cases of Ge, Ar and NaI detectors, which are reported
by the COHERENT collaboration to start measuring CEvNS in the near
future. Table~\ref{Tab:01} gives information about the estimated mass,
threshold and baseline on each case, all of which will be considered
in our following computations to predict the current estimated number
of events by using Eq.~(\ref{SM_Events}).

\section{Sensitivity to the Weak mixing angle}
Future CEvNS measurements will determine with accuracy the weak mixing
angle value.  Any deviation from the Standard Model
prediction~\cite{Tanabashi:2018oca,Erler:2017knj} for this important
quantity will be an indicator of new physics. Although the current
estimates for the weak mixing angle from the CsI measurement are not
competitive~\cite{Kosmas:2017tsq}, future information from CEvNS may
be of important relevance for this test of the SM at very low
energies~\cite{Cadeddu:2018izq,Huang:2019ene}. For example, this
information can be useful for the atomic parity
  violation (APV) measurement, where a small deviation from the
prediction has been found~\cite{Cadeddu:2018izq}.  As already
mentioned, we have studied the future sensitivity to the weak mixing
angle for the next generation of COHERENT
experiments~\cite{Akimov:2018ghi}. To this purpose, we have assumed
that a futuristic $\chi^{2}$ analysis will be given by the
minimization of the function:

\begin{equation}
\chi ^{2} = \left (\frac{N^{exp} - (1+\alpha)N^{th}(X) - (1+\beta)N^{bg}}{\sigma}  \right )^{2} + \left ( \frac{\alpha}{\sigma_{\alpha}} \right )^{2} + \left ( \frac{\beta}{\sigma_{\beta}} \right )^{2},
\label{chi_sq}
\end{equation}
\newline
where $N^{exp}$ is the measured number of events, which, as we are
dealing with a future experiment, we will consider as given by the SM
prediction plus the expected background, $N^{th}(X)$ represents the
predicted number of events as a function of a set of variable
parameters $X$, which in this case corresponds only to the weak mixing
angle, $N^{bg}$ is the expected background number of events, that we
will consider to be ten percent of the predicted number of events;
this background could come, for instance, from neutrino induced
neutrons~\cite{Kolbe:2000np} or prompt neutrons~\cite{Akimov:2017ade}.
This general expression will also be used in the following sections.
The statistical uncertainty is given by $\sigma = \sqrt{N^{exp}}$, and
the parameters $\alpha$ and $\beta$ quantify the systematic errors
with an associated uncertainty $\sigma_{\alpha,\beta}$.

\begin{figure}[ht] 
\begin{center}
\includegraphics[width=0.3\textwidth]{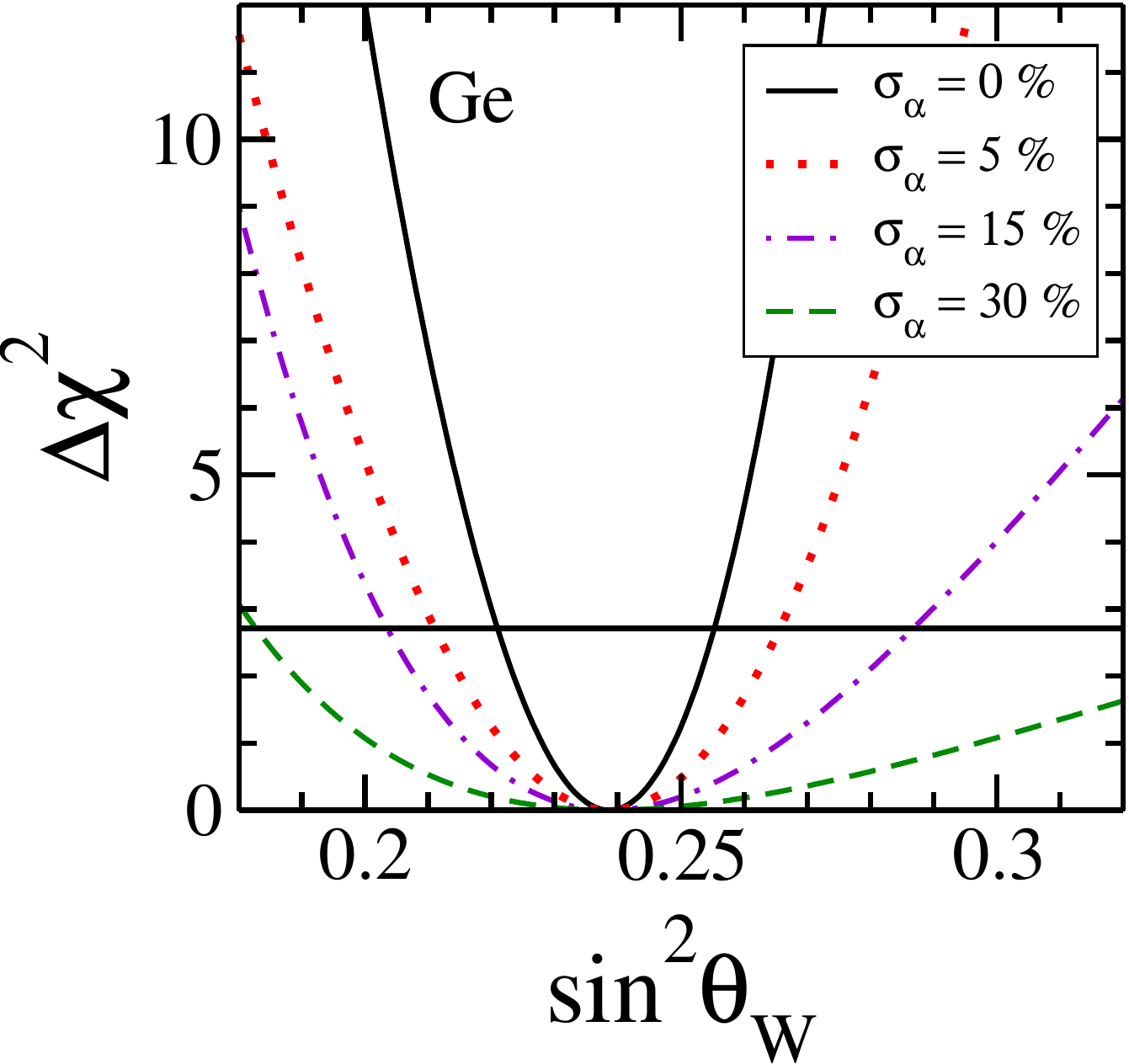}
\includegraphics[width=0.3\textwidth]{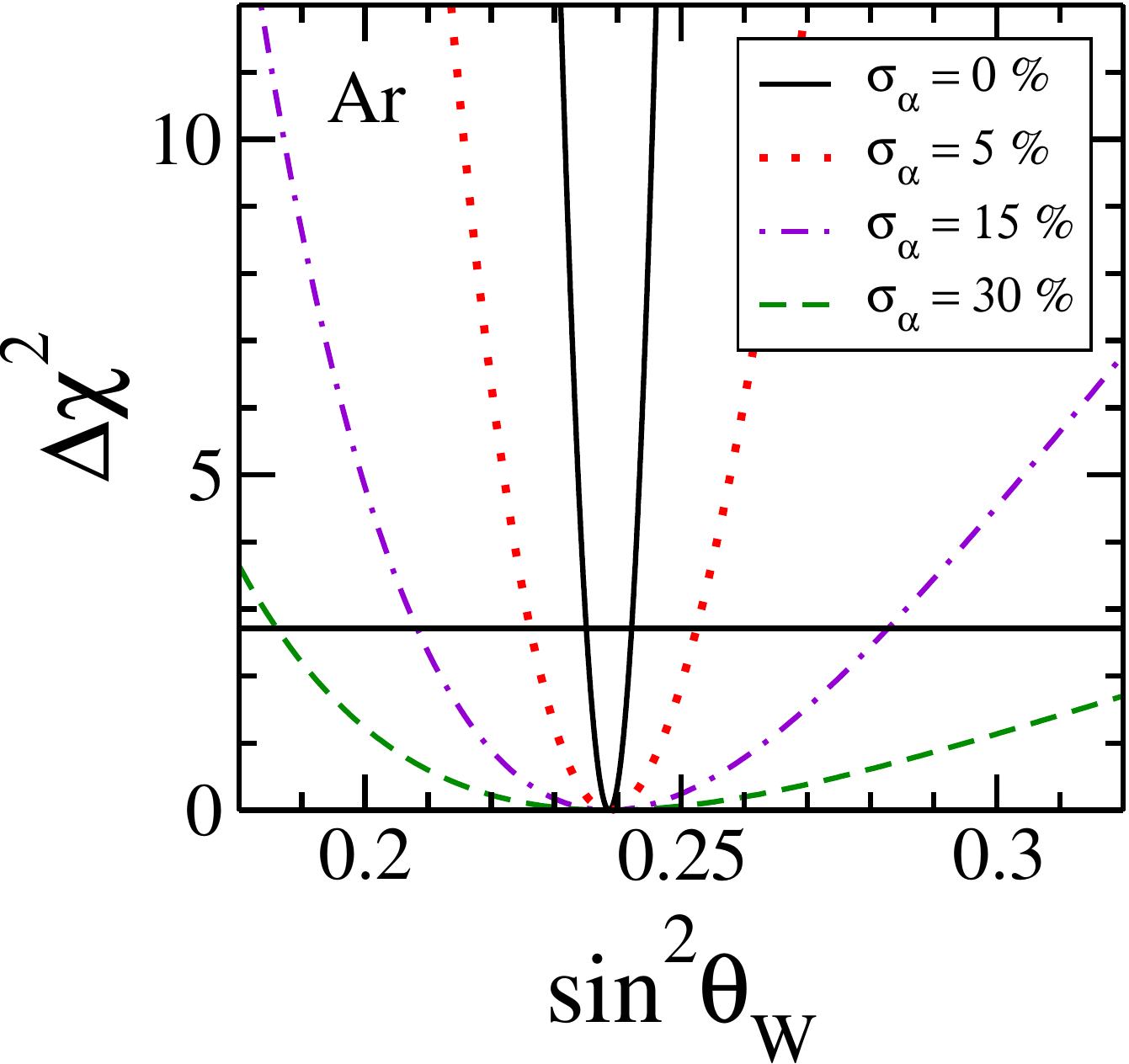}
\includegraphics[width=0.3\textwidth]{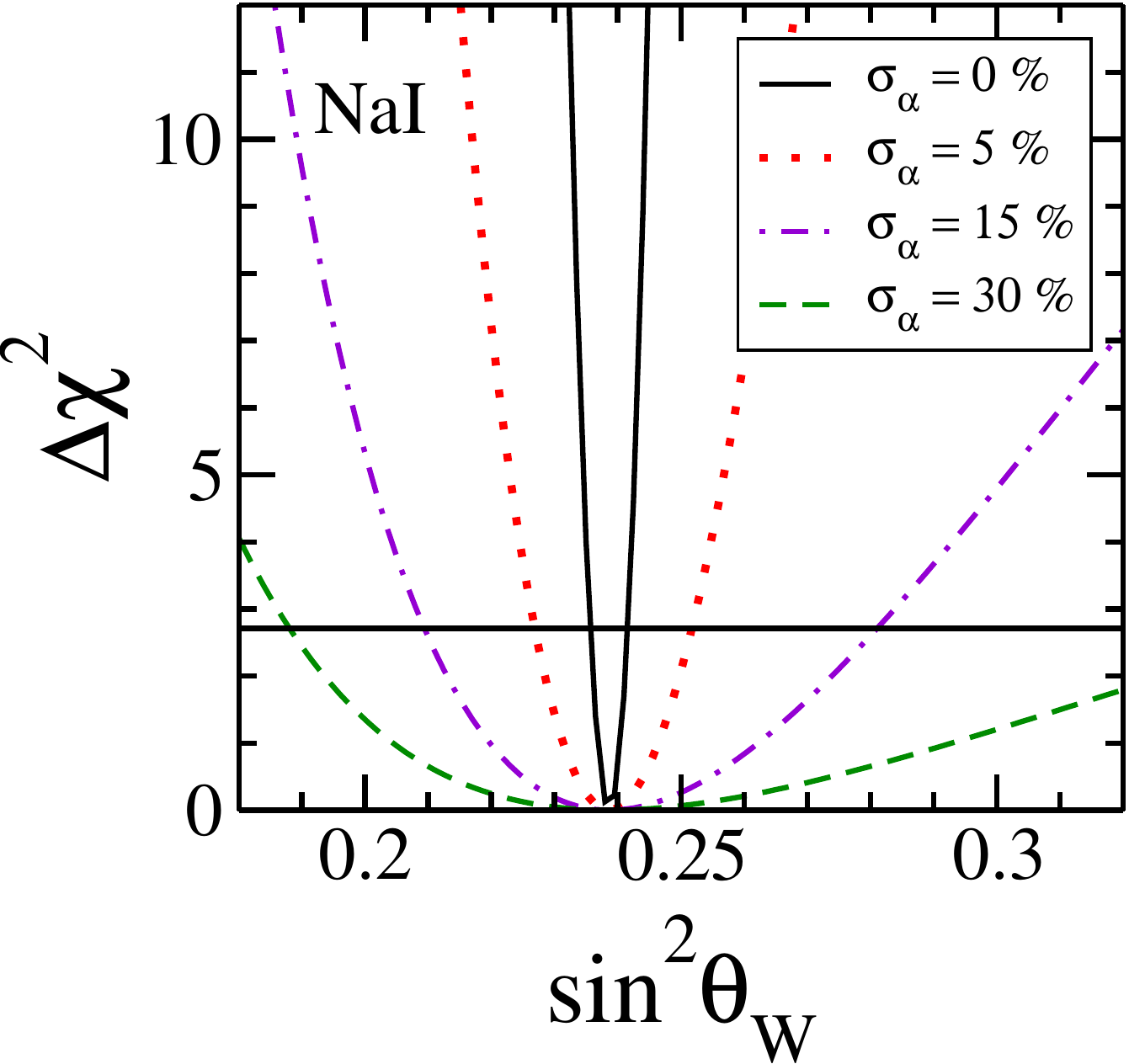}
\end{center}
\caption{\label{fig:wma} Expected sensitivity to
  $\sin^2\theta_W$ for the different detectors under consideration:
  Germanium, Argon, and NaI, respectively. The different curves are
  for the ideal case of $\sigma_\alpha = 0$~\% 
with $100$~\% efficiency and a background error of $\sigma_\beta = 0$~\%  
(solid),
for $\sigma_\alpha = 5$~\% 
with $50$~\% efficiency and a background error of $\sigma_\beta = 10$~\%  
(dotted),  
for $\sigma_\alpha = 15$~\% 
with $100$~\% efficiency and a background error of $\sigma_\beta = 10$~\% 
(dashed-dotted),
and $\sigma_\alpha = 30$~\% 
with $100$~\% efficiency and a background error of $\sigma_\beta = 10$~\%  
(dashed), see Table~\ref{Tab:03} and text for details. The horizontal line indicates 
the $90$~\% CL. }
\end{figure}

The results are shown in Table~\ref{Tab:03} and Fig.~\ref{fig:wma},
where we have taken four different scenarios for each detector. These
scenarios were considered in order to illustrate what would be the
CEvNS sensitivity to this parameter.  Regarding the systematic error
$\sigma_\alpha$, we can estimate that a first measurement could have
large errors due to the normalization and quenching factor among other
systematics~\cite{Akimov:2017ade}. Therefore we have considered a
first scenario with $\sigma_\alpha=30$~\%, which is similar to the one
reported by the first measurement reported by the COHERENT
collaboration. A second case is that of $\sigma_\alpha=15$~\% that
might be a realistic one after the detectors are better
characterized. For the case of the background error we have considered
it as $10$~\%. We also show as a reference the very ideal case of no
errors. Finally a different scenario with $50$~\% efficiency and
$\sigma_\alpha=5$~\%, $\sigma_\beta=10$~\% is also considered. The
expected sensitivity for the weak mixing angle (for symmetrized
errors) is shown in Table~\ref{Tab:03}, where we reported it at a
$90$~\% CL.  With these different scenarios, we expect to present a
broad idea of the possible constraints that future experiments can
obtain and what type of error would be more important to control. In
all our computations we have considered an acceptance function equal
to the unity for all $T$.

We can see from Table~\ref{Tab:03} and
Fig.~\ref{fig:wma} that, as expected, a detector with larger mass, such
as the $NaI$ case, will give better constraints, provided that the
systematic errors are under control. In any case, even the more modest
case of the Germanium detector with a $10$~kg array could give a
competitive measurement (for low energies) if the systematics can be
maintained under control. 

\begin{table}
  \begin{tabular}{l c c c c c c c  } \hline \hline 
Experiment    &  $50$ \%  eff &  & $100$~\%  eff  & & $100$~\% eff  
& & $100$~\% eff \\ 
             &  $\sigma_{\alpha} = 5$~\%  &  & $\sigma_{\alpha}=0$~\%  eff  & & $\sigma_{\alpha} = 15$~\% 
& & $\sigma_{\alpha} = 30$~\%  \\ 
             &  $\sigma_{\beta} = 10$~\%  &  & $\sigma_{\beta}=0$~\%  eff  & & $\sigma_{\beta} = 10$~\% 
& & $\sigma_{\beta} = 10$~\%  \\ 
  \hline \hline
Ge       & 11.5 & & 7.2  & &  17.5 & & 36.8 \\
Ar       & 5.5 & & 1.5  & &  15.6 & & 35.6 \\
NaI      & 5.2 & & 1.2  & &  15.0 & & 34.2 \\ 
\hline \hline 
\end{tabular}\caption{\label{Tab:03} Expected sensitivity, in percent, to the weak mixing angle. For each experiment we show the 
$90$~\% expected sensitivity for the different cases that we have considered. 
}
\end{table}

\section{Sensitivity to NSI}
Besides the precision tests of the Standard Model, there has been a
lot of interest in different extensions of the SM to explain, for
instance, the neutrino mass pattern. A useful phenomenological approach
is that of non-standard interactions
(NSI)~\cite{Farzan:2017xzy,Miranda:2015dra,Ohlsson:2012kf}.  In
general, neutral current non-standard interactions can be parametrized
by introducing a Lagrangian of the form:

\begin{equation}
\mathfrak{L} _{\nu H}^{NSI}=-\frac{G_{F}}{\sqrt{2}} \sum_{\substack{q=u,d,\\\alpha,\beta = e,\mu,\tau}}\left [\overline{\nu} _{\alpha }\gamma ^{\mu }\left ( 1-\gamma ^{5} \right )\nu _{\beta } \right ]\left ( \varepsilon _{\alpha \beta }^{qL}\left [\overline{q}\gamma _{\mu }\left ( 1-\gamma ^{5} \right )q \right ]+\varepsilon _{\alpha \beta }^{qR}\left [\overline{q}\gamma _{\mu }\left ( 1+\gamma ^{5} \right )q \right ] \right ).
\label{NS}
\end{equation}

Here the interaction is modeled between the neutrino and the $up$ and
$down$ quarks within the nucleons, so the index $q$ runs over $u$ and
$d$. The subscripts $\alpha$ and $\beta$ run over the three flavors
$e, \mu$, and $\tau$. The Lagrangian in Eq.~(\ref{NS}) contains flavor
preserving, non-universal, non-standard terms which are proportional
to $\varepsilon _{\alpha \alpha}^{qV}$ (with $V = L, R$).  Also, it
contains the so-called flavor-changing terms proportional to
$\varepsilon _{\alpha \beta}^{qV}$ with $\alpha \neq \beta$; all these
coupling constants are taken in terms of the Fermi constant. Thus, for
an electron (anti)neutrino source, the cross-section for $T <<
E_{\nu}$ now
reads~\cite{Barranco:2005yy,Scholberg:2005qs,AristizabalSierra:2018eqm,Dent:2017mpr,Lindner:2016wff,Coloma:2017egw}:

\begin{equation}
\begin{aligned}
\frac{\mathrm{d} \sigma }{\mathrm{d} T}\left ( E_{\nu },T \right ) \simeq& \frac{G_{F}^{2}M}{\pi }\left ( 1-\frac{MT}{2E_{\nu }^{2}} \right )\left \{ \left [ Z\left ( g_{V}^{p}+2\varepsilon _{ee}^{uV}+\varepsilon _{ee}^{dV} \right )F_{Z}^{V}(Q^{2})+N\left ( g_{V}^{n}+\varepsilon _{ee}^{uV}+2\varepsilon _{ee}^{dV}  \right ) F_{N}^{V}(Q^{2})\right ]^{2}\right.\\
&\left. +\sum_{\alpha }\left [ Z\left ( 2\varepsilon _{\alpha e}^{uV}+\varepsilon _{\alpha e}^{dV} \right )F_{Z}^{V}(Q^{2})+N\left ( \varepsilon _{\alpha e}^{uV}+2\varepsilon _{\alpha e}^{dV} \right )F_{Z}^{V}(Q^{2}) \right ]^{2} \right \} .
\end{aligned}
\label{CrossN}
\end{equation}

\begin{figure}[ht] 
\begin{center}
\includegraphics[width=0.3\textwidth]{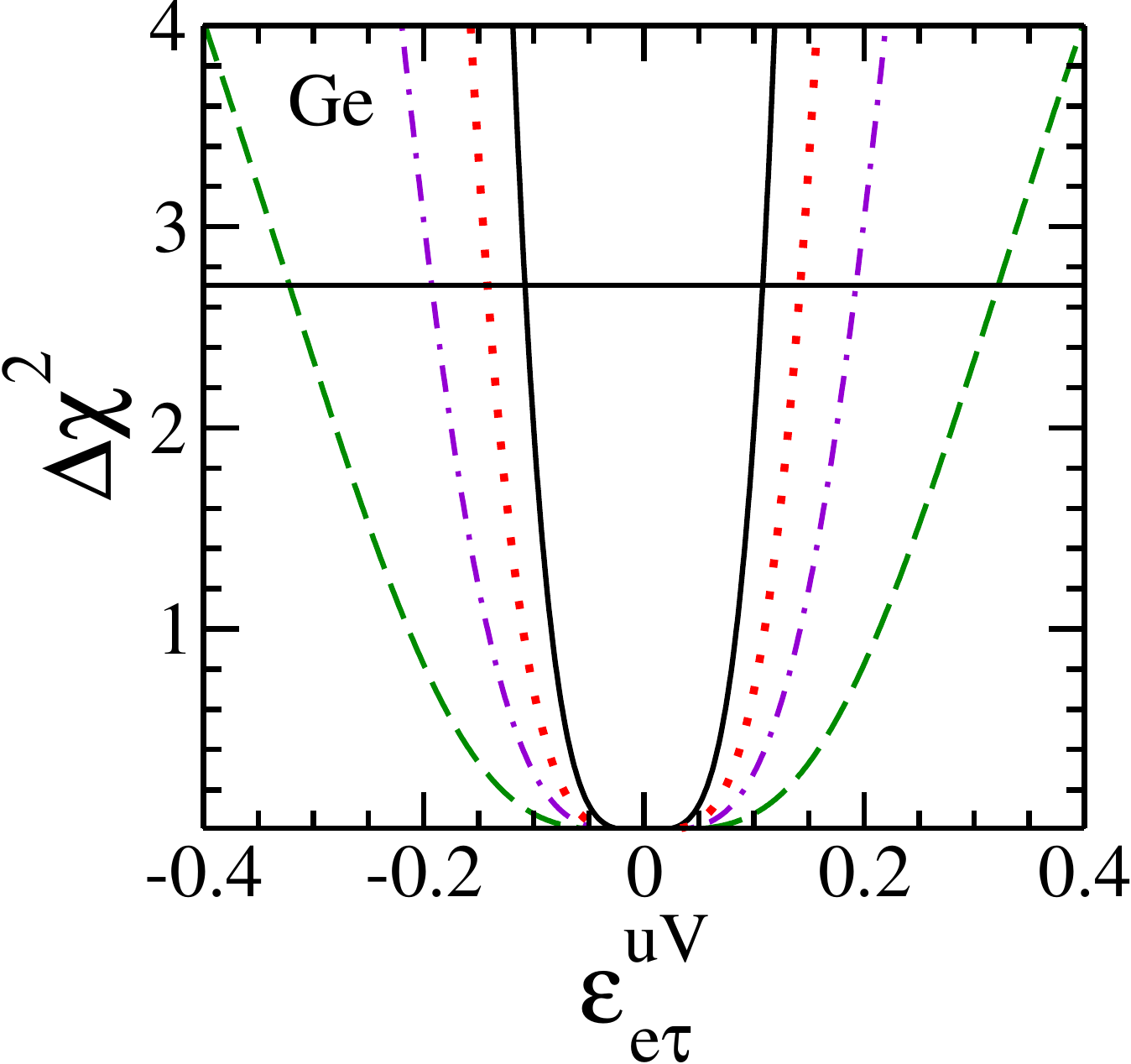}
\includegraphics[width=0.3\textwidth]{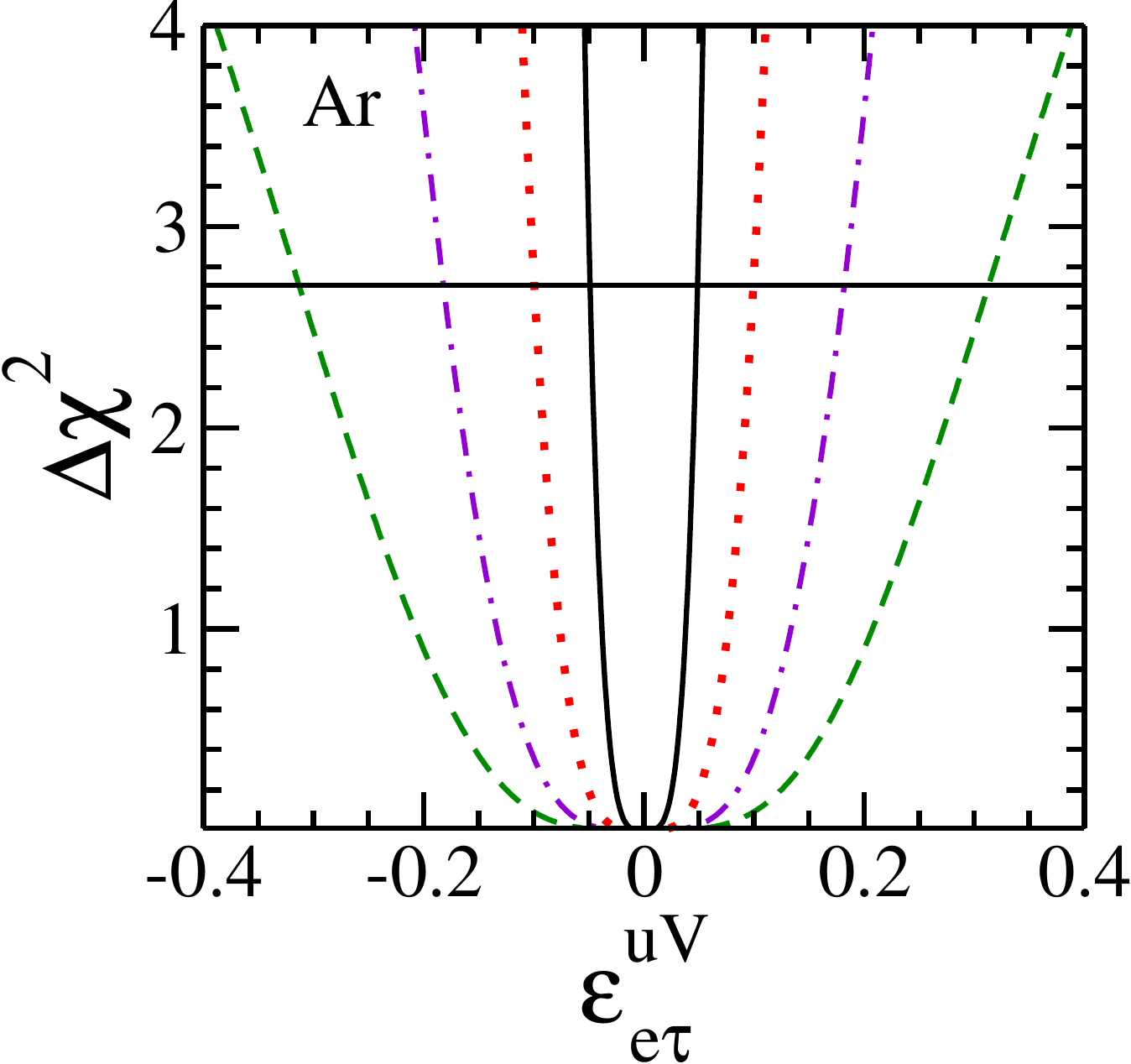}
\includegraphics[width=0.3\textwidth]{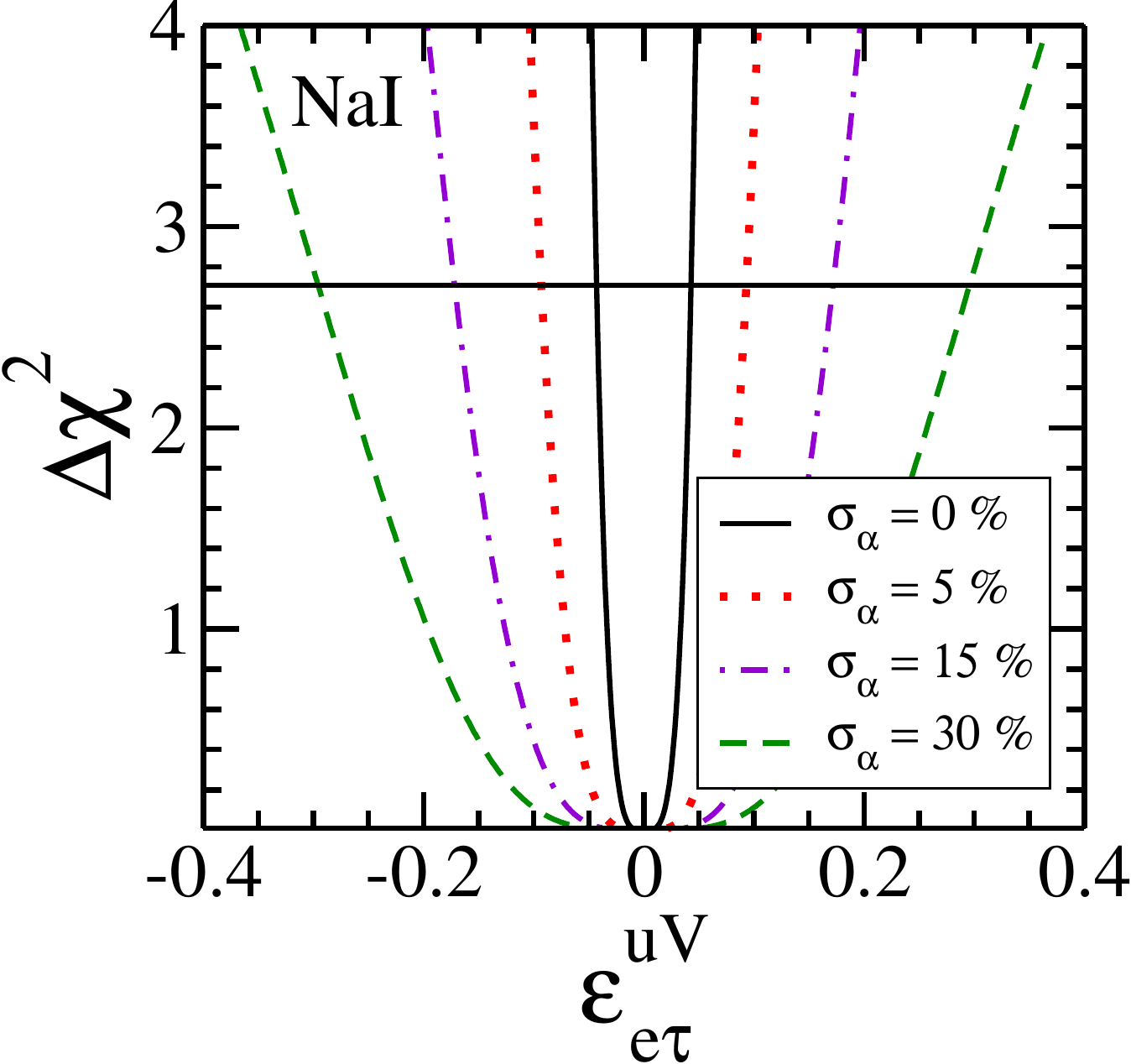}
\end{center}
\caption{\label{fig:FC-NSI} 
Expected sensitivity to
  $\epsilon^{uV}_{e\tau}$ for the different detectors under consideration:
  Germanium, Argon, and NaI, respectively. 
Again, 
the different curves are
  for the ideal case of $\sigma_\alpha = 0$~\% 
with $100$~\% efficiency and a background error of $\sigma_\beta = 0$~\%  
(solid),
for $\sigma_\alpha = 5$~\% 
with $50$~\% efficiency and a background error of $\sigma_\beta = 10$~\%  
(dotted),  
for $\sigma_\alpha = 15$~\% 
with $100$~\% efficiency and a background error of $\sigma_\beta = 10$~\% 
(dashed-dotted),
and $\sigma_\alpha = 30$~\% 
with $100$~\% efficiency and a background error of $\sigma_\beta = 10$~\%  
(dashed), see text for details. The horizontal line indicates 
the $90$~\% CL. 
}
\end{figure}
\begin{table}
  \begin{tabular}{l c c c c c c c  } \hline \hline 
Experiment    &  $50$ \%  eff &  & $100$~\%  eff  & & $100$~\% eff
& & $100$~\%  eff \\ 
             &  $\sigma_{\alpha} = 5$~\%  &  & $\sigma_{\alpha}=0$~\%  eff  & & $\sigma_{\alpha} = 15$~\% 
& & $\sigma_{\alpha} = 30$~\%  \\ 
             &  $\sigma_{\beta} = 10$~\%  &  & $\sigma_{\beta}=0$~\%  eff  & & $\sigma_{\beta} = 10$~\% 
& & $\sigma_{\beta} = 10$~\%  \\ 
  \hline \hline
Ge       & $|\varepsilon_{\tau e}^{uV}| < 0.142$ & & 
 $|\varepsilon_{\tau e}^{uV}| < 0.108$  & & 
 $|\varepsilon_{\tau e}^{uV}| < 0.193$  & & 
 $|\varepsilon_{\tau e}^{uV}| < 0.322$ \\
Ar       & $|\varepsilon_{\tau e}^{uV}| < 0.100$ & & 
 $|\varepsilon_{\tau e}^{uV}| < 0.048$ & &  
 $|\varepsilon_{\tau e}^{uV}| < 0.182$ & & 
 $|\varepsilon_{\tau e}^{uV}| < 0.314$ \\
NaI      & $|\varepsilon_{\tau e}^{uV}| <  0.093$ & & 
 $|\varepsilon_{\tau e}^{uV}| < 0.041$ & &  
 $|\varepsilon_{\tau e}^{uV}| < 0.172$ & & 
 $|\varepsilon_{\tau e}^{uV}| < 0.296$  \\ 
\hline \hline 
\end{tabular}\caption{\label{Tab:02} Expected sensitivity to the 
flavor changing NSI parameter $\varepsilon_{\tau e}^{uV}$. For each
experiment we quote the $90$~\%~CL 
expected sensitivity for the different scenarios that we have considered. 
}
\end{table}

\begin{figure}[ht] 
\begin{center}
\includegraphics[width=0.3\textwidth]{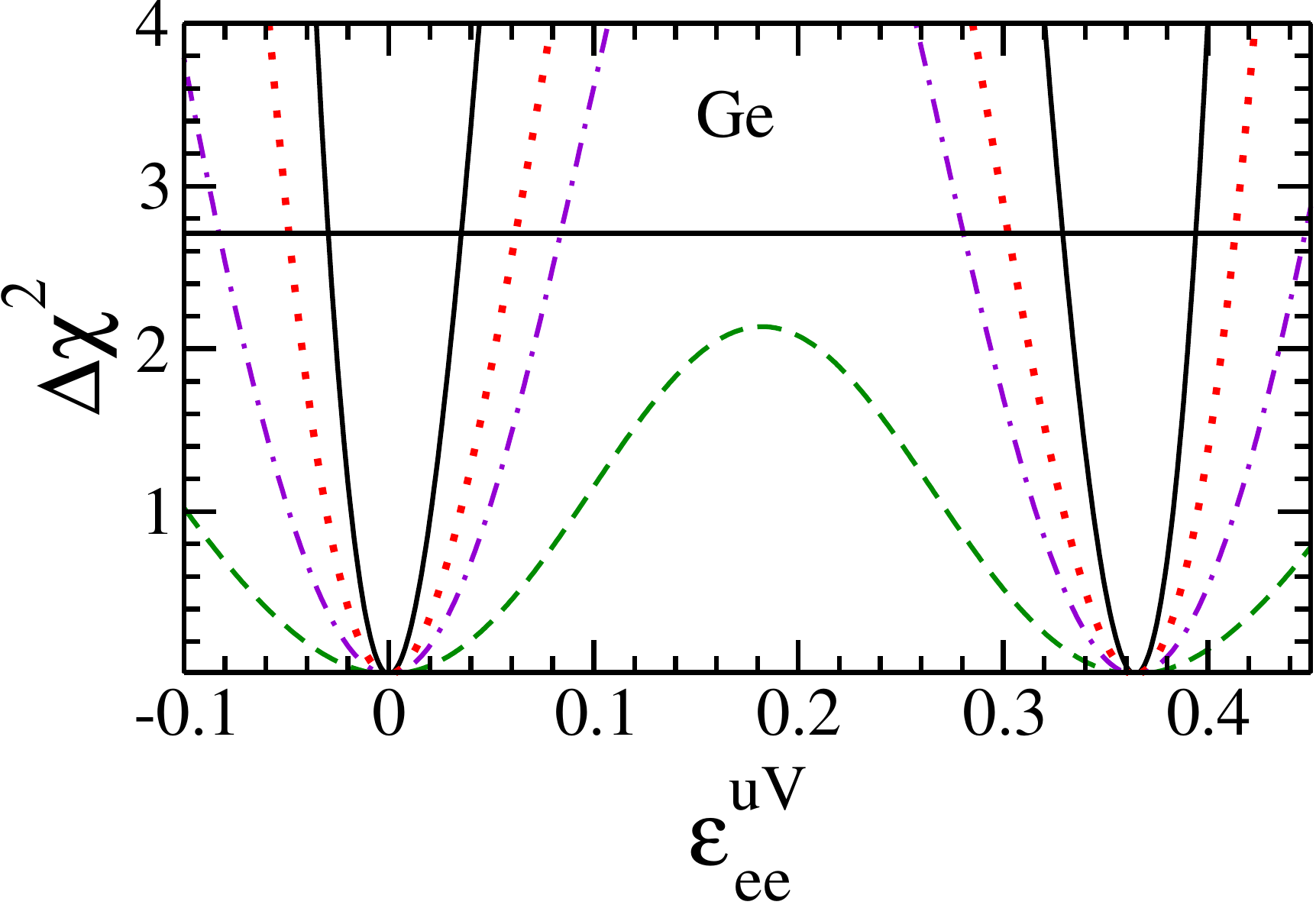}
\includegraphics[width=0.3\textwidth]{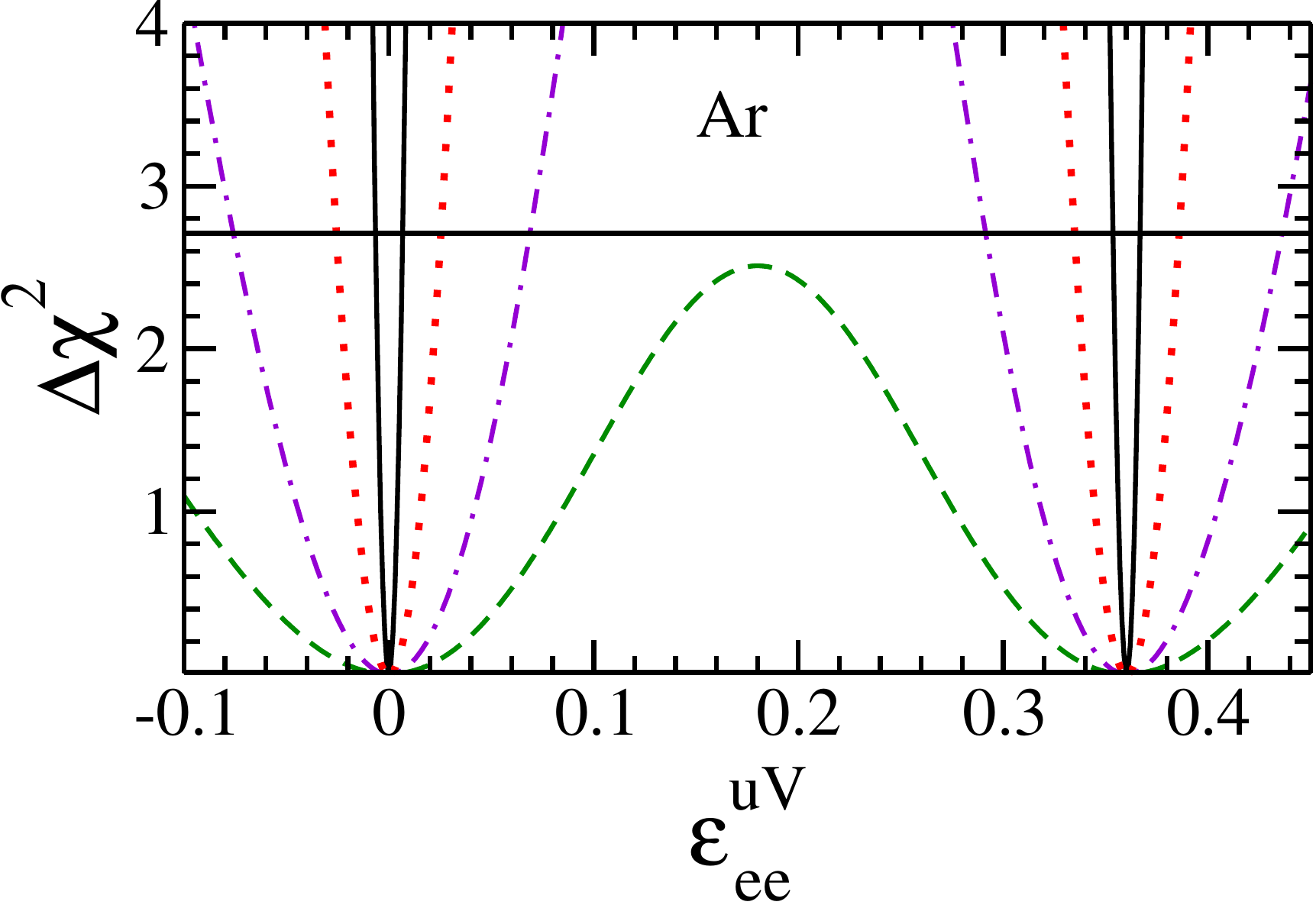}
\includegraphics[width=0.3\textwidth]{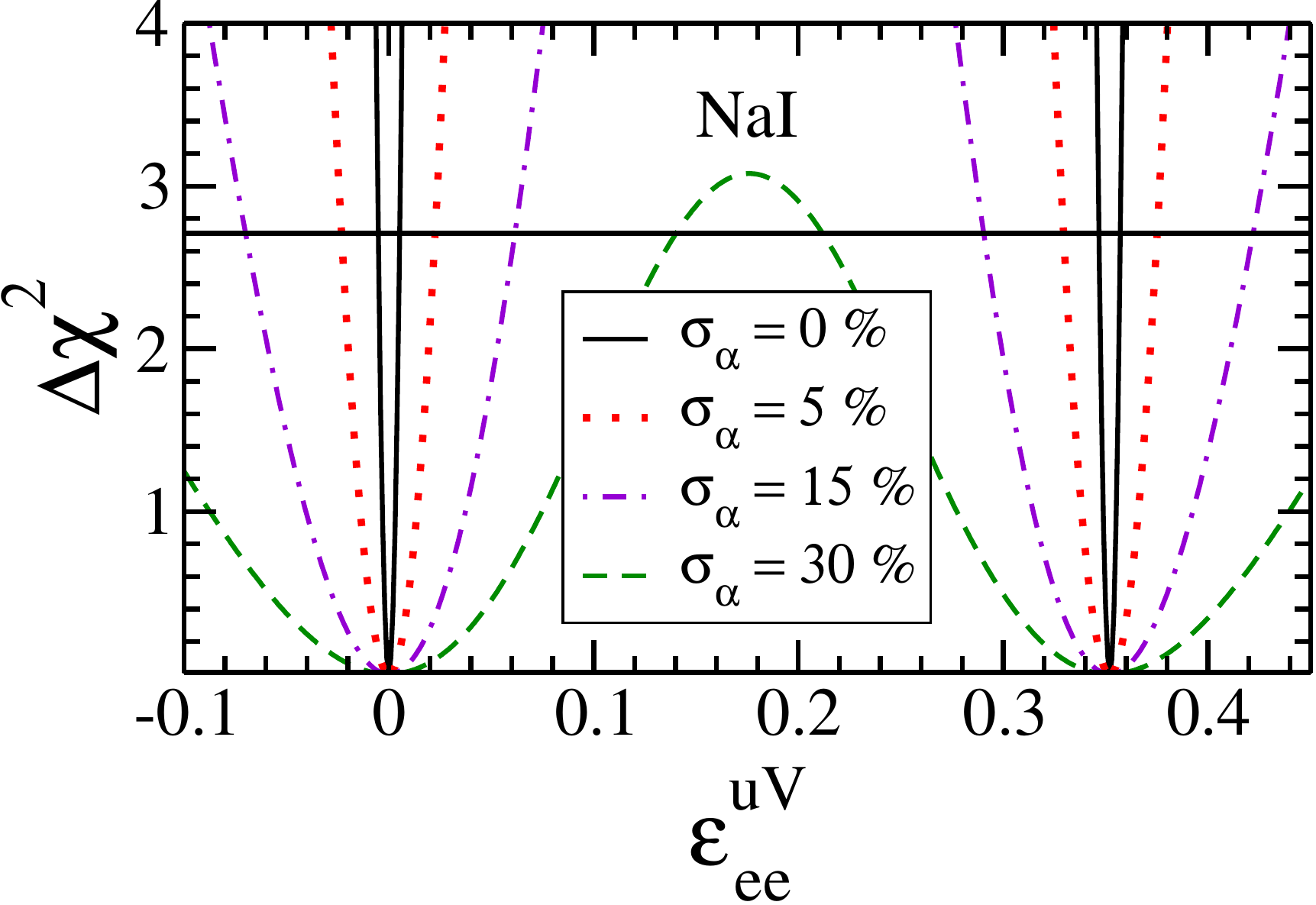}
\end{center}
\caption{\label{fig:NU-NSI} Expected sensitivity to
  $\epsilon^{uV}_{ee}$ for the different detectors under
  consideration: Germanium, Argon, and NaI, respectively. As in
  previous cases, 
the different curves are
  for the ideal case of $\sigma_\alpha = 0$~\% 
with $100$~\% efficiency and a background error of $\sigma_\beta = 0$~\%  
(solid),
for $\sigma_\alpha = 5$~\% 
with $50$~\% efficiency and a background error of $\sigma_\beta = 10$~\%  
(dotted),  
for $\sigma_\alpha = 15$~\% 
with $100$~\% efficiency and a background error of $\sigma_\beta = 10$~\% 
(dashed-dotted),
and $\sigma_\alpha = 30$~\% 
with $100$~\% efficiency and a background error of $\sigma_\beta = 10$~\%  
(dashed), see text for details. The horizontal line indicates 
the $90$~\% CL. }
\end{figure}

The cross section for a muon (anti)neutrino source has the same form
and can be obtained by exchanging the indices $e \leftrightarrow \mu$.
For simplicity, here we will only consider the possibility of having
NSI interactions coming from the electron neutrino source; this is a
natural choice since the muon NSI parameters are usually more
restricted from other
experiments~\cite{Farzan:2017xzy,Miranda:2015dra,Ohlsson:2012kf}.
This means that the number of events measured by a detector at the SNS
will be given by:

\begin{equation}
N^{th} = N_{D}\int_{T}A(T)dT\int_{E_{min}}^{52.8 MeV}dE \sum_{a} \frac{\mathrm{d}N_{a}}{\mathrm{d}E}\frac{\mathrm{d} \sigma_{a} }{\mathrm{d} T},
\label{N_NSI}
\end{equation}
\newline
where $a = \bar{\nu}_{\mu}, \nu_{\mu}, \nu_{e}$, with $\frac{\mathrm{d}
  \sigma_{a} }{\mathrm{d} T}$ given by Eq.~(\ref{CrossN}) for $a = \nu_{e}$
and by Eq.~(\ref{eq:00}) for the other two cases.

The study of the sensitivity to the NSI parameters is
of relevance since any positive signal will hint for new physics; on
the other hand, constraints on these parameters will potentially
discard models of new physics. This is the case, for instance, for the
first measurement of CEvNS where the reported
constraints~\cite{Akimov:2017ade,Coloma:2017ncl,Liao:2017uzy,Kosmas:2017tsq}
disfavored a class of
models~\cite{Farzan:2015doa,Farzan:2015hkd} that were
motivated by the Dark-LMA
solution~\cite{Miranda:2004nb,Escrihuela:2009up,Coloma:2016gei,Gonzalez-Garcia:2013usa}.  Future constraints
from COHERENT collaboration will allow to set stronger constraints on
the NSI parameters. The use of intense neutrino sources close to a
CEvNS detector allows for a powerful setup that strongly constraints
NSI parameters, competitive with any other neutrino experiment as
already pointed out in Refs.~\cite{Barranco:2005yy,Scholberg:2005qs}

As in the previous section, we have studied the potential of the
future setups for the SNS and computed the expected sensitivity for
the different future detectors that are to be installed.  We first
made the analysis considering only one NSI parameter to be non-zero at
a time. Although the parameters are correlated, this will give a first
idea of the constraints that can be obtained and could be useful when
we expect only small deviations from zero from a given new physics
model~\cite{Valle:2015pba,Fukugita:2003en}. We will end up this
section with a two-parameters analysis.  Again, we have computed the
$\chi^{2}$ analysis of Eq.~(\ref{chi_sq}) with $N^{th}$ given by
Eq.~(\ref{N_NSI}), and $X$ representing the corresponding NSI
parameter. This time we have also considered the four different
scenarios for the futuristic systematic uncertainties as in the
previous section.
Fig.~\ref{fig:FC-NSI} shows the
results for $\varepsilon _{\tau e}^{uV}$, while
Fig.~\ref{fig:NU-NSI} shows the corresponding results for
$\varepsilon _{ee}^{uV}$. We can notice that, for the case of
non-universal parameters, there are two different intervals where the
$\varepsilon _{ee}^{uV}$ values can lie. This is a well-known
degeneracy that appears in the CEvNS case~\cite{Scholberg:2005qs}.  We
show the numerical restrictions for both the flavor-changing and
non-universal parameters in Table~\ref{Tab:02} and Table~\ref{Tab:04}, respectively.
As in the previous section, we can see the importance of 
systematic errors, efficiency, and the mass of the detector. For
example, we can notice that for the case of a two tons NaI detector,
the sensitivity is such that, even with a $30$~\% error, the
experiment can tell between the two degenerate allowed regions for
$\varepsilon _{ee}^{uV}$, as can be seen in Table~\ref{Tab:04}.

\begin{table}
\begin{tabular}{l c c c c c c c } \hline \hline 
Experiment    &  $50$~\%  eff &  & $100$~\%  eff  & & $100$~\% eff 
& & $100$~\%  eff\\ 
             &  $\sigma_{\alpha} = 5$~\%  &  & $\sigma_{\alpha}=0$~\%  eff  & & $\sigma_{\alpha} = 15$~\% 
& & $\sigma_{\alpha} = 30$~\%  \\ 
             &  $\sigma_{\beta} = 10$~\%  &  & $\sigma_{\beta}=0$~\%  eff  & & $\sigma_{\beta} = 10$~\% 
& & $\sigma_{\beta} = 10$~\%  \\ 
\hline \hline
Ge & -0.049 $<\varepsilon_{ee}^{uV}<$ 0.062 & & 
-0.030 $<\varepsilon_{ee}^{uV}<$ 0.035 & & 
-0.083 $<\varepsilon_{ee}^{uV}<$ 0.084 & & 
-0.188 $<\varepsilon_{ee}^{uV}<$ 0.553\\ 
   & 0.302 $<\varepsilon_{ee}^{uV}<$ 0.414 & & 
0.328 $<\varepsilon_{ee}^{uV}<$ 0.394& & 
0.28 $<\varepsilon_{ee}^{uV}<$ 0.445& & \\
Ar & -0.026 $<\varepsilon_{ee}^{uV}<$ 0.026 & & 
-0.006 $<\varepsilon_{ee}^{uV}<$ 0.007  & & 
-0.076 $<\varepsilon_{ee}^{uV}<$ 0.069  & & 
-0.182 $<\varepsilon_{ee}^{uV}<$ 0.542 \\
   & 0.335 $<\varepsilon_{ee}^{uV}<$ 0.386& & 
0.353 $<\varepsilon_{ee}^{uV}<$ 0.366 & & 
0.291 $<\varepsilon_{ee}^{uV}<$ 0.437& &  \\
NaI & -0.023 $<\varepsilon_{ee}^{uV}<$ 0.023 & & 
-0.004 $<\varepsilon_{ee}^{uV}<$ 0.004 & & 
-0.070 $<\varepsilon_{ee}^{uV}<$ 0.062 & & 
-0.169 $<\varepsilon_{ee}^{uV}<$ 0.141 \\ 
    & 0.329 $<\varepsilon_{ee}^{uV}<$ 0.375& & 
0.347 $<\varepsilon_{ee}^{uV}<$ 0.356& & 
0.290 $<\varepsilon_{ee}^{uV}<$ 0.422& & 
0.211 $<\varepsilon_{ee}^{uV}<$ 0.521\\ 
\hline \hline 
\end{tabular}\caption{\label{Tab:04} Expected sensitivity to the 
non-universal NSI parameter $\varepsilon_{ee}^{uV}$. For each
experiment we quote the $90$~\%~CL 
expected sensitivity for the different scenarios that we have considered. 
}
\end{table}

\begin{figure}[ht] 
\begin{center}
\includegraphics[width=0.3\textwidth]{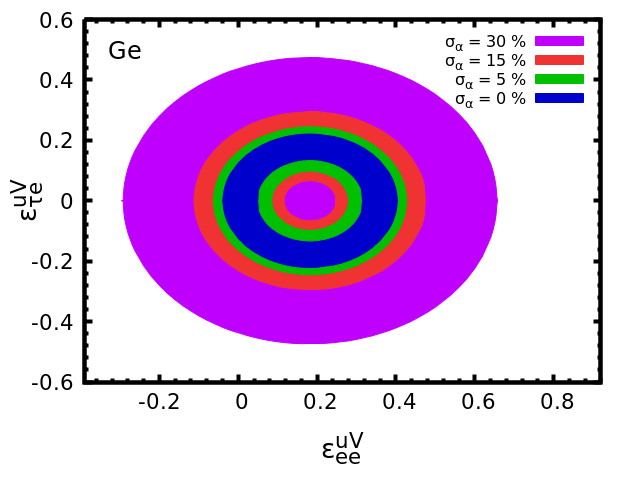}
\includegraphics[width=0.3\textwidth]{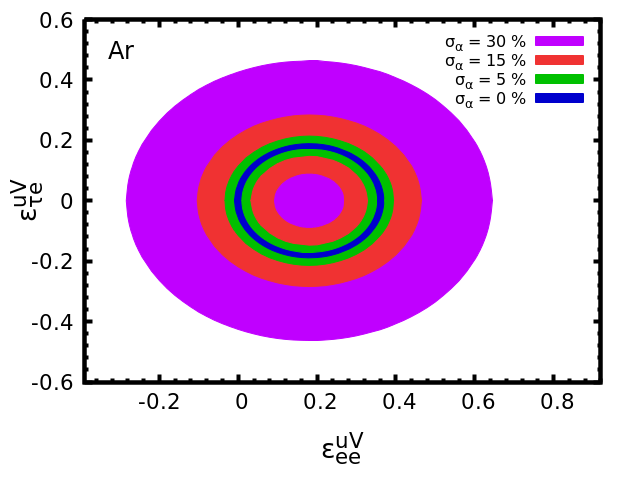}
\includegraphics[width=0.3\textwidth]{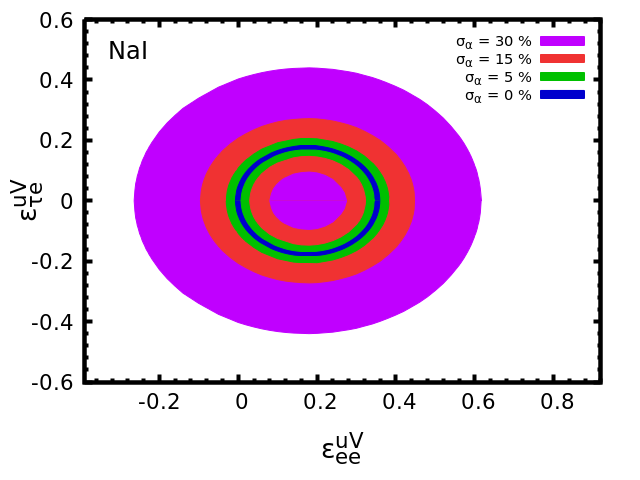}
\end{center}
\caption{\label{fig:2P-NSI} Expected sensitivity to
    $\epsilon^{uV}_{ee}$ vs $\epsilon^{uV}_{e\tau}$, at $90$~\% CL for
    the different detectors under consideration: Germanium, Argon, and
    NaI, respectively. As in previous cases, the different curves are
    for the ideal case of $\sigma_\alpha = 0$~\% with $100$~\%
    efficiency and a background error of $\sigma_\beta = 0$~\% (blue),
    for $\sigma_\alpha = 5$~\% with $50$~\% efficiency and a
    background error of $\sigma_\beta = 10$~\% (green), for
    $\sigma_\alpha = 15$~\% with $100$~\% efficiency and a background
    error of $\sigma_\beta = 10$~\% (red), and $\sigma_\alpha = 30$~\%
    with $100$~\% efficiency and a background error of $\sigma_\beta =
    10$~\% (magenta), see text for details. } 
\end{figure}

We close this section by illustrating one of the degeneracies that can
appear if we consider more than one NSI parameter
different from zero~\cite{Barranco:2005yy,Scholberg:2005qs}.  For the
case of considering both $\epsilon_{ee}^{uV}$ and
$\epsilon_{e\tau}^{uV}$ different from zero, we can notice from
Eq.~(\ref{CrossN}) that there is a degeneration in the determination
of this parameters that give rise to an ellipse, in analogy with
Ref.~\cite{Barranco:2005yy}.  We illustrate this case in
Fig.~(\ref{fig:2P-NSI}), where we can see that for this specific combination 
of parameters the detectors are unable to remove the degeneracy. 

\section{Sensitivity to the sterile neutrino hypothesis}
Currently, the three neutrino oscillation picture is well established
and most of its parameters are well
measured~\cite{deSalas:2017kay,Capozzi:2018ubv,Esteban:2018azc}. However,
there are different neutrino flux anomalies that cannot be explained
by considering neutrino oscillations between three neutrino
flavors~\cite{Gariazzo:2017fdh}. For instance, the LSND observes an
appearance of a $\bar{\nu_e}$ on a $\bar{\nu_\mu}$ flux, MIniBoone
measures an excess of $\nu_e$ and $\bar{\nu_e}$ that agrees with the
LSND results. On the other hand, for electron antineutrinos, a
disappearance of $\bar{\nu_e}$ is observed in experiments with reactor
neutrinos. These effects may be explained by considering a fourth
non-interacting, sterile neutrino flavor. Expected constraints,
considering different experimental setups, have been considered for
the CEvNS case~\cite{Dutta:2015nlo,Kosmas:2017zbh,Canas:2017umu}.

By considering each neutrino flavor state as a linear combination of
mass eigenstates

\begin{equation}
    \nu_l=\sum\limits_{m}U_{lm} \nu_m
\end{equation}
\newline
where U is a unitary mixing matrix, we can find the oscillation
probability to be given by~\cite{Giunti:2007ry}:

\begin{equation}
    P_{\nu_\alpha\nu_\beta}=\delta_{\alpha\beta}-4\sum\limits_{i>j}Re(U^*_{\alpha i}U_{\beta i}U_{\alpha j}U^*_{\beta j})\sin^2\left(\Delta m^2_{ij}\frac{L}{4E} \right)+2\sum\limits_{i>j}Im(U^*_{\alpha i}U_{\beta i}U_{\alpha j}U^*_{\beta j})\sin\left(\Delta m^2_{ij}\frac{L}{2E} \right)
\end{equation}
\newline
where the latin subscripts correspond to the mass eigenstates, and the
greek ones correspond to the $e,\mu,\tau,s$ neutrino flavors, $\Delta
m_{ij}^2=m_i^2-m_j^2$, $L$ is the distance from the neutrino source to
the detector, and $E_\nu$ is the neutrino energy. Assuming CPT
invariance, the anti-neutrino case is found by interchanging each
matrix element with its complex conjugate, resulting in a reverse of
the signs.

Due to the short source to detector distance ($\sim 10$~m) and to the
SNS neutrino energy spectrum ($\sim 10$~MeV), the oscillations between
the three active states can be neglected. Therefore, the probability
of oscillation from active to sterile states can be studied in a
two-flavor approximation:

\begin{equation}
\label{eq:probSterile}
    P_{\nu_\alpha\nu_s}=\sin^22\theta_{\alpha\beta}\sin^2\left(\frac{1.27\Delta m^2_{i4}L}{E_\nu} \right) .
\end{equation}
\newline
For simplicity, we will consider two different cases. First, the
oscillation from $\nu_e\rightarrow \nu_s$ and then the corresponding
case for muon (anti)neutrinos.  To take into account the oscillation
of the neutrinos produced at the SNS, we take the probability that the
considered neutrino keeps the same flavor as:

\begin{equation}
    P_{\alpha}=1-\sin^22\theta_{\alpha\alpha}\sin^2\left(\frac{1.27\Delta m^2_{i4}L}{E_\nu} \right)
\end{equation}
this oscillation probabilty is multiplied by the neutrino flux and
integrated over the neutrino energy spectrum, so the total number of
events expected is given in this two cases by:

\begin{equation}
N^{th} = N_{D}\int_{T}A(T)dT\int_{E_{min}}^{52.8 MeV}dE \sum_{\alpha}\frac{dN_{\alpha}}{dE}P_\alpha(\theta_{\alpha\alpha},\Delta m^2_{i4})\frac{\mathrm{d} \sigma }{\mathrm{d} T}, \label{eq:NthOsc}
\end{equation}

\begin{figure}[ht] 
\begin{center}
\includegraphics[width=0.3\textwidth]{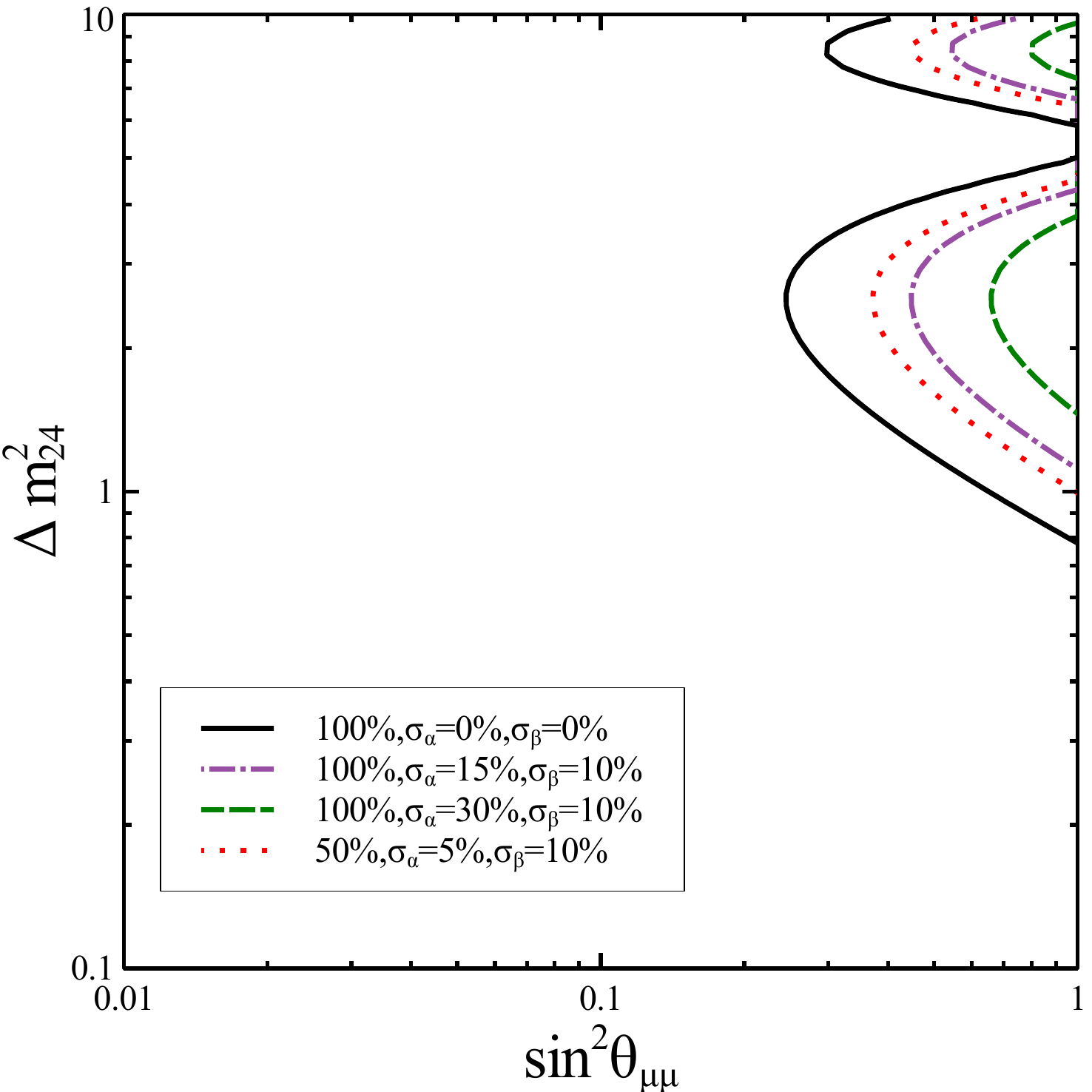}
\includegraphics[width=0.3\textwidth]{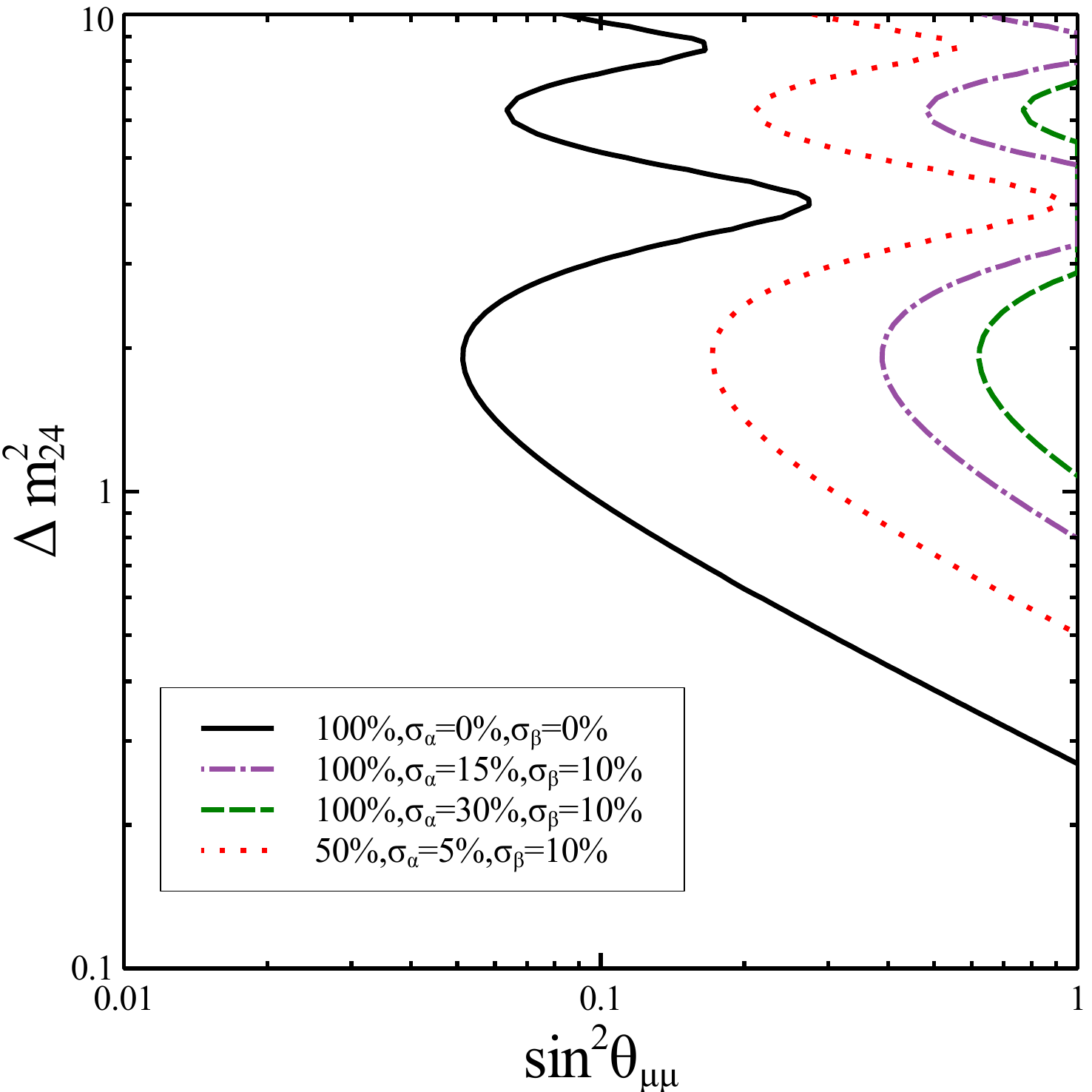}
\includegraphics[width=0.3\textwidth]{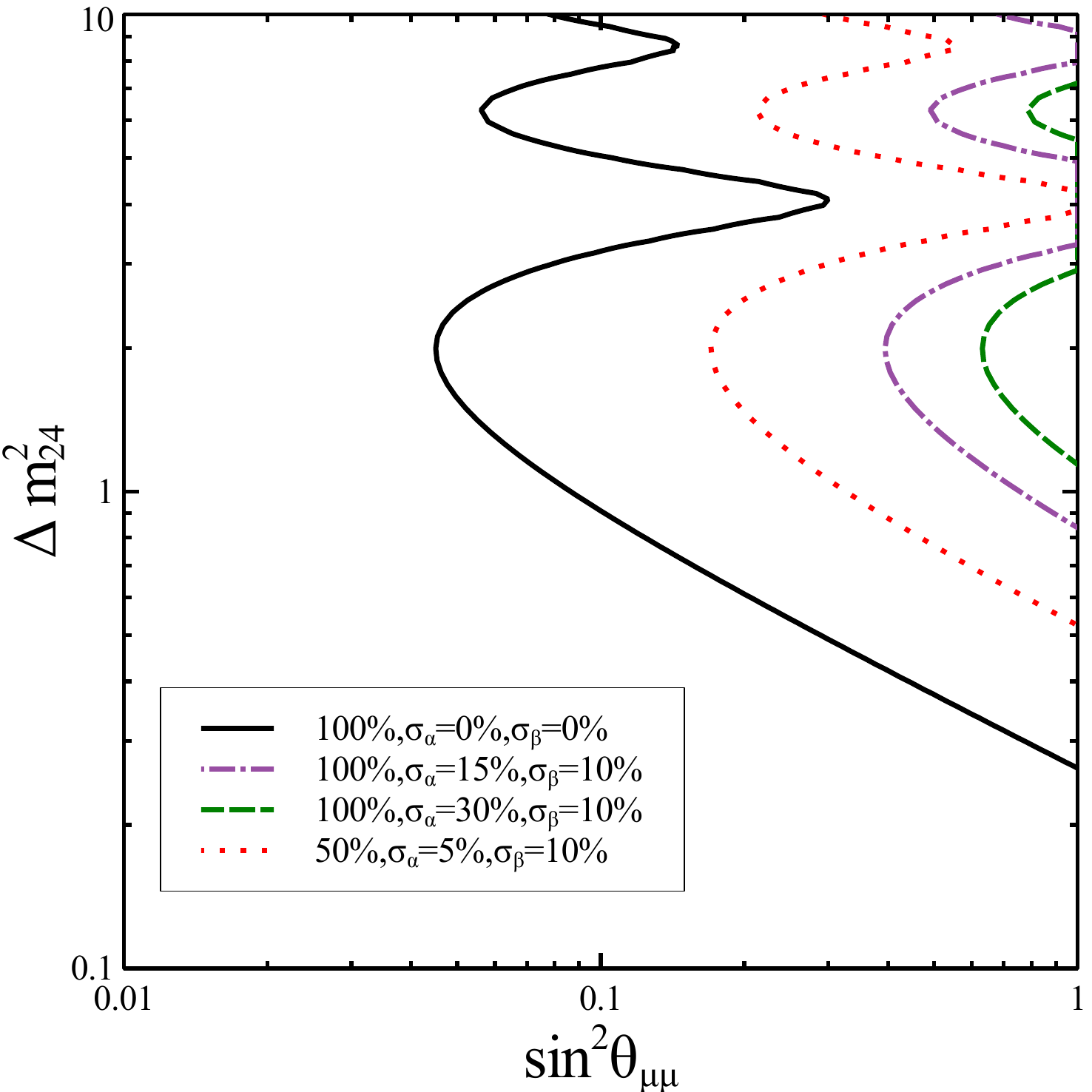}
\end{center}
\caption{\label{fig:sterile24} 
Expected sensitivity for a muon neutrino oscillation into a sterile neutrino 
  state, for the different detectors under consideration:
Germanium (left), Argon (middle), and NaI (right), respectively. 
Again, 
the different curves are
  for the ideal case of $\sigma_\alpha = 0$~\% 
with $100$~\% efficiency and a background error of $\sigma_\beta = 0$~\%  
(solid),
for $\sigma_\alpha = 5$~\% 
with $50$~\% efficiency and a background error of $\sigma_\beta = 10$~\%  
(dotted),  
for $\sigma_\alpha = 15$~\% 
with $100$~\% efficiency and a background error of $\sigma_\beta = 10$~\% 
(dashed-dotted),
and $\sigma_\alpha = 30$~\% 
with $100$~\% efficiency and a background error of $\sigma_\beta = 10$~\%  
(dashed), see text for details. The horizontal line indicates 
the $90$~\% CL. 
}
\end{figure}

\begin{figure}[ht] 
\begin{center}
\includegraphics[width=0.3\textwidth]{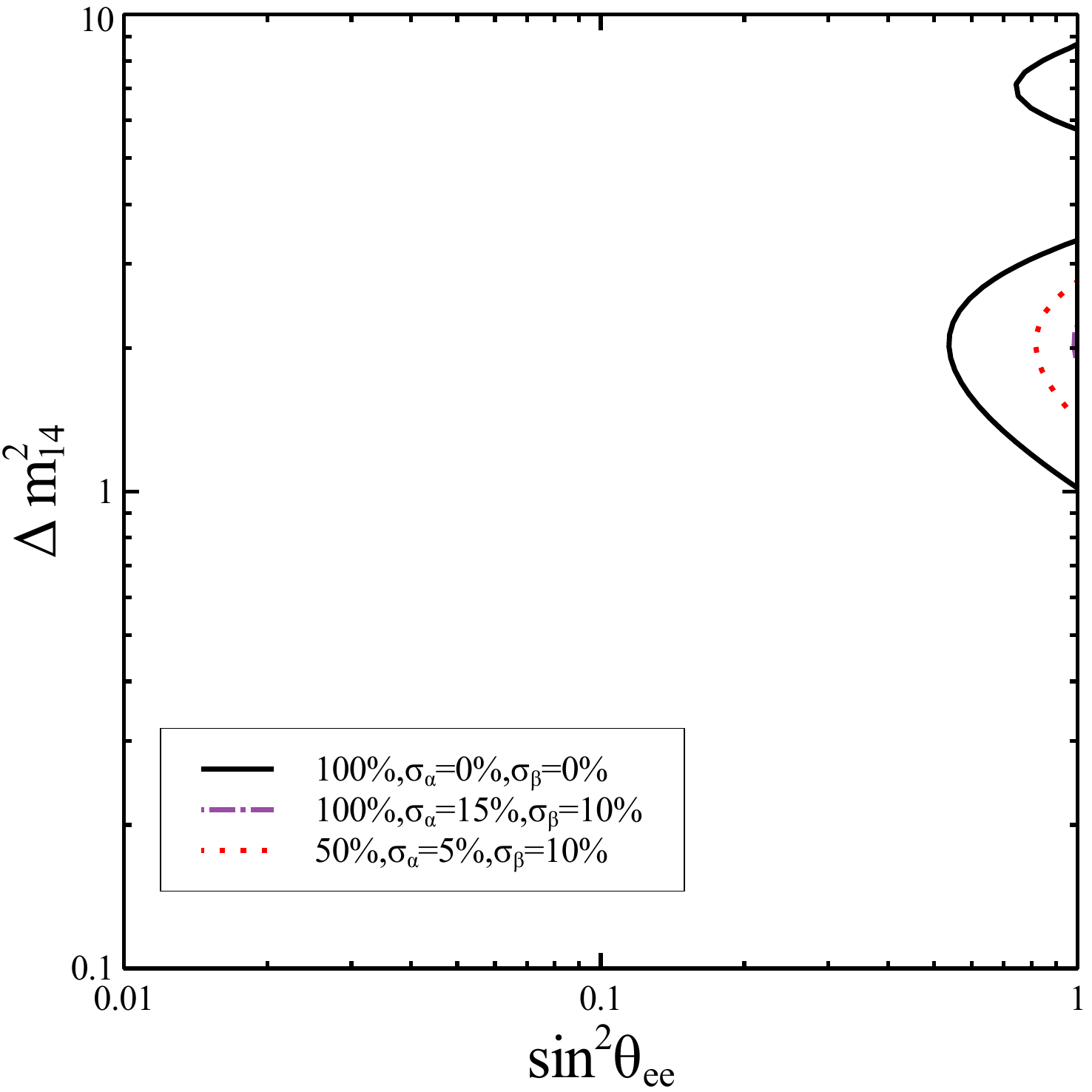}
\includegraphics[width=0.3\textwidth]{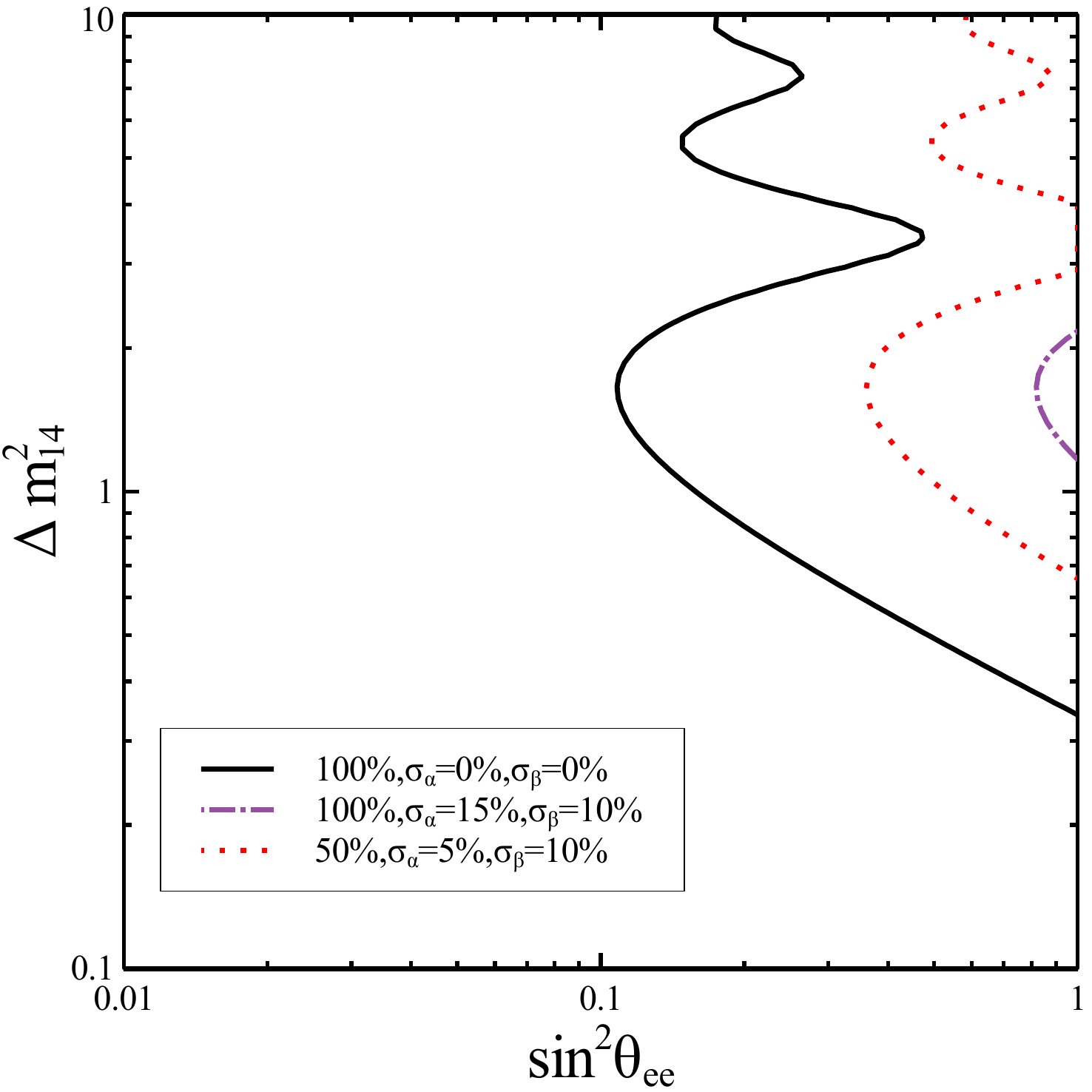}
\includegraphics[width=0.3\textwidth]{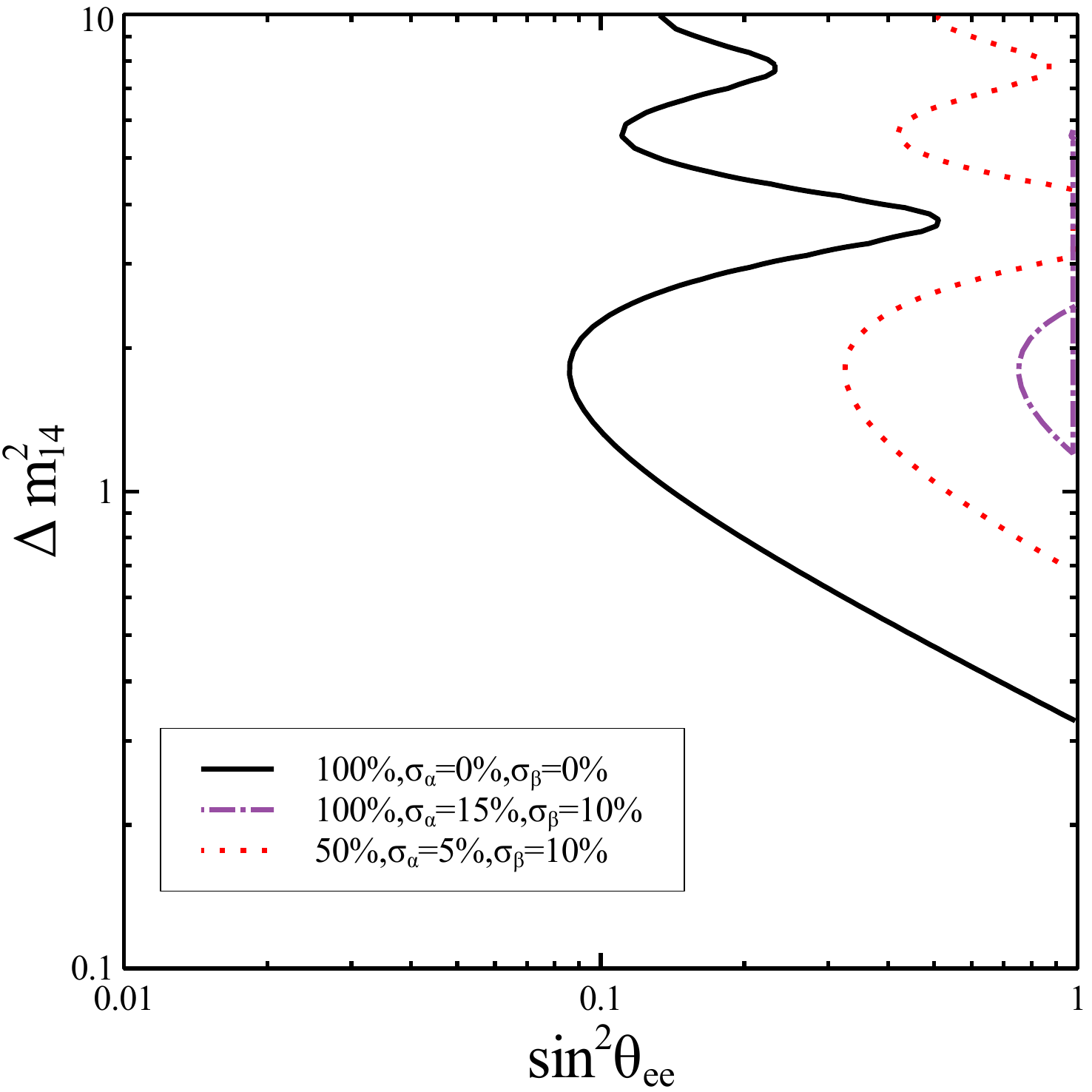}
\end{center}
\caption{\label{fig:sterile14} Expected sensitivity for an electron
  neutrino oscillation into a sterile neutrino state, for the
  different detectors under consideration: Germanium (left), Argon
  (middle), and NaI (right), respectively. 
Again, 
the different curves are
  for the ideal case of $\sigma_\alpha = 0$~\% 
with $100$~\% efficiency and a background error of $\sigma_\beta = 0$~\%  
(solid),
for $\sigma_\alpha = 5$~\% 
with $50$~\% efficiency and a background error of $\sigma_\beta = 10$~\%  
(dotted),  
for $\sigma_\alpha = 15$~\% 
with $100$~\% efficiency and a background error of $\sigma_\beta = 10$~\% 
(dashed-dotted),
and $\sigma_\alpha = 30$~\% 
with $100$~\% efficiency and a background error of $\sigma_\beta = 10$~\%  
(dashed), see text for details. The horizontal line indicates 
the $90$~\% CL. 
}
\end{figure}

where the fluxes for the different neutrino flavors, $\alpha$, are
defined by Eqs.~(\ref{FluxDelta}-\ref{FluxElectron}). As in the case
of NSI, we considered a $\chi^2$ function in order to forecast the
sensitivity of COHERENT future experiments. In this case, since
neutrino oscillation probability is a function of two variables (
$\sin^22\theta_{\alpha\alpha},\Delta m^2_{i4}$), we will take the $\chi^2$ function 
as the one described in Eq.~(\ref{chi_sq}) with $N^{th}(X) = N^{th}(\sin^22\theta_{\alpha\alpha},\Delta m^2_{i4})$ as in Eq.~(\ref{eq:NthOsc}). 

For this case, our analysis considers only one parameter at a
time. That is, we only consider either $\sin\theta_{ee}$ or
$\sin\theta_{\mu\mu}$ different from zero and compute the corresponding
effect in the electron (muon) neutrino number of events. In
Figs.~\ref{fig:sterile24} and~\ref{fig:sterile14} we show the sensitivity
to the allowed regions of the parameters
$\sin^22\theta_{\alpha\alpha}$ and $\Delta m^2_{i4}$ for the 
different systematic errors and efficiencies that we have already
discussed. The results are shown at $90$~\% CL.

Although for some cases the expected sensitivity is not competitive,
we can notice that for the more ambitious detectors with larger mass
there is sensitivity to the relevant region of sterile neutrino searches. 

\section{Discussion and conclusions}
The measurement of CEvNS by the COHERENT collaboration has been a
break through that opens the door to new measurements of this ellusive
process. Motivated by the future program of the same collaboration, we
have studied the expected sensitivity for precision tests of the
Standard Models as well as for new physics searches. We have focused
on the case of the measurement of the weak mixing angle, the
sensitivity to NSI and the future constraints on a sterile neutrino
state.

We have studied the different proposed detectors on equal footing, in
the sense that we have considered the same total neutrino flux coming
from the spallation neutron source and we have also considered the same
efficiencies and systematic errors. We have illustrated quantitatively
that the most ambitious large mass detector arrays will give better
constraints on new physics, provided that systematics are under
control. We have also estimated the weakness of the constraints if the
efficiency is compromised.

\acknowledgments{This work was supported by CONACYT-Mexico under grant
  A1-S-23238 and by SNI (Sistema Nacional de Investigadores). OGM
  would also like to thank COFI. The 
  data used to support the findings of this study are available from
  the corresponding author upon request. }

\end{document}